\documentclass[aps,pra,superscriptaddress,reprint,amsmath,amssymb]{revtex4-2}
\usepackage{graphicx}

\newif\ifstructure
\structuretrue

\newcommand{\beq}{\begin{equation}}
\newcommand{\eeq}{\end{equation}}
\newcommand{\bea}{\begin{eqnarray}}
\newcommand{\eea}{\end{eqnarray}}

\newcommand{\cm}{cm$^{-1}$}

\usepackage{afterpage}
\usepackage{empheq}

\usepackage{soul}

\usepackage[caption=false]{subfig}
\usepackage{float}

\usepackage{xcolor}

\usepackage{multirow}     % Para fusionar celdas
\usepackage{array}        % Para ajustar columnas

\usepackage{longtable}

\usepackage{appendix}

\begin{document}

\title{Molecular structure, binding, and disorder in TDBC–Ag plexcitonic assemblies}
\author{J. \surname{Baños-Gutiérrez}}
\affiliation{Instituto de Quimica, Universidad Nacional Autonoma de Mexico, Circuito Exterior, Ciudad Universitaria, Alcaldía Coyoacán C.P. 04510, Ciudad de Mexico}
\author{R. \surname{Bercy}}
\affiliation{Institute for Integrative Biology of the Cell, CEA, CNRS, Université Paris Saclay, CEA Saclay 91191 Gif sur Yvette Cedex, France.}
\author{Y. \surname{García Jomaso}}
\affiliation{Instituto de Física, Universidad Nacional Autónoma de México, Apartado Postal 20-364, Ciudad de México, C.P. 01000, Mexico.}
\author{S. \surname{Balci}}
\affiliation{Department of Photonics, Izmir Institute of Technology, 35430 Izmir, Turkey}
\author{G. \surname{Pirruccio}}
\affiliation{Instituto de Física, Universidad Nacional Autónoma de México, Apartado Postal 20-364, Ciudad de México, C.P. 01000, Mexico.}
\author{J. \surname{Halldin Stenlid}}
\affiliation{Department of Chemistry and Chemical Engineering, Chalmers University of Technology, Kemiv\"{a}gen 10, SE-412 96, Gothenburg, Sweden}
\author{M.J. \surname{Llansola-Portoles}}
\affiliation{Institute for Integrative Biology of the Cell, CEA, CNRS, Université Paris Saclay, CEA Saclay 91191 Gif sur Yvette Cedex, France.}
\author{D. \surname{Finkelstein-Shapiro}}
\email{daniel.finkelstein@iquimica.unam.mx}
\affiliation{Instituto de Quimica, Universidad Nacional Autonoma de Mexico, Circuito Exterior, Ciudad Universitaria, Alcaldía Coyoacán C.P. 04510, Ciudad de Mexico}

\begin{abstract}
Plexcitonic assemblies are hybrid materials composed of a plasmonic nanoparticle and molecular or semiconducting emitters whose electronic transitions are strongly coupled to the plasmonic mode. This coupling hybridizes the system modes into upper and lower polariton branches. The strength of the interaction depends on the number of emitters and on their orientation and spatial arrangement relative to the metallic surface. These structural factors have profound consequences for the ensuing photoexcited dynamics.
Despite the extensive spectroscopic work on plexcitonic systems, direct understanding of the molecular geometry at the metal interface remains limited. In this work, we present a comprehensive structural characterization of one of the most widely studied plexcitons—formed by the cyanine dye 5,5',6,6'-tetrachloro-1,1'-diethyl-3,3'-di(4-sulfobutyl)-benzimidazolocarbocyanine (TDBC) and silver nanoprisms—using a combination of NMR, THz-Raman spectroscopy, and DFT calculations. By comparing the signals from the monomeric and aggregated forms of TDBC with that of the plexciton, we identify shared spectral fingerprints that reveal how molecular packing is modified when the aggregate adsorbs on the silver surface. We observe Raman modes specific to plexciton systems, and identify NOESY cross-peaks in the aliphatic region, that along with THz-Raman modes in the 10-400 \cm region are sensitive indicators of aggregation geometry and adsorption. 
We find that isolated TDBC monomers adopt an asymmetric conformation in which both sulfobutyl chains lie on the same side of the chromophore, while J-aggregates adopt a symmetric up–down alternation of the chains from molecule to molecule. This work establishes the molecular geometry of a prototypical TDBC–silver plexciton, providing a structural benchmark for understanding geometry–dependent photophysics in hybrid exciton–plasmon systems. \end{abstract}

\maketitle

\section{Introduction}

%General paragraoh on plexcitons
Plexcitons are composed of a plasmonic nanoparticle decorated by semiconducting or molecular emitters whose radiative transition strongly couples to the plasmonic mode, and forms hybrid light-matter, or polaritonic  states \cite{Hertzog2019,Manuel2019,Bitton2022,Peruffo2023,Bitton2019,Torma2015,Zengin2015}. These states are strongly delocalized, and have been proposed for long distance energy transport \cite{Zhong2016,Schachenmayer2015,Feist2015,Engelhardt2022}, lasing \cite{Kena-Cohen2010}
%,Deng2003} XXX
, and chemical reactivity \cite{Martinez-Martinez2018a,Ribeiro2018,xiang2024molecular}. 
% ,Ribeiro2018,Herrera2016}. XXX
% 
% Importance of the orientation 
The orientation and distance between molecule and metal influences the strength of the coupling and the energetic landscape in which the ultrafast dynamics occurs. 
In contrast to the behavior observed in Fabry-Perot based polaritons, where molecules lie in the middle of the cavity, far away from the mirrors, the molecules in plexcitons exist at the surface of the nanoparticle, opening interfacial charge and energy transfer channels \cite{YuenZhou2025,Finkelstein2021}. 
The most promising plexcitonic systems use highly anisotropic nanoparticles such as prisms \cite{Balci2016}, urchins \cite{Peruffo2023,Peruffo2021,nicola_peruffo_engineering_2022}, donuts, and disks \cite{balci2019colloidal}, thereby creating a corresponding anisotropy in the metal-molecule coupling \cite{Yankovich2019}.

Disorder plays a nontrivial role in cavity polaritons by mediating the hybridization between bright collective excitations and the dark excitonic manifold. In disordered molecular ensembles, energetic and orientational disorder mixes dark excitons with the cavity-coupled bright mode, thereby reshaping the effective non-Hermitian polaritonic Hamiltonian. Using an exact Green’s-function treatment of disordered quantum emitters in multimode cavities, Engelhardt and Cao showed that disorder renormalizes polariton dispersion and produces a crossover from coherent underdamped dynamics to overdamped relaxation as disorder increases \cite{Engelhardt2022}. Importantly, this framework demonstrates that dark states actively participate in determining the cavity local density of states and relaxation pathways rather than acting as a passive reservoir. Extending beyond single-mode and one-dimensional models, Sun et al. demonstrated that in realistic three-dimensional multimode Fabry–Pérot cavities, dark states can become highly delocalized across hundreds of molecules, with a participation ratio that scales linearly with the number of molecules in the plane parallel to the cavity mirrors, even in the presence of energetic and orientational disorder. Together, these results establish that disorder does not simply localize polaritons, but instead controls the extent, geometry, and dimensionality of dark-state delocalization, with direct implications for energy transport, relaxation bottlenecks, and cavity-modified chemistry \cite{Sun2025}. A central difficulty in applying these theoretical frameworks to plexcitons is the lack of direct experimental knowledge of molecular geometry and structural disorder.

In plexcitonic systems, where molecular excitons hybridize with localized surface plasmon resonances, disorder plays an even more central role due to the intrinsically open, lossy, and spatially inhomogeneous nature of plasmonic nanostructures \cite{Schirato2023}. Structural disorder in molecular aggregates (e.g., energetic inhomogeneity, finite exciton delocalization length, and polymorphism) coexists with electromagnetic heterogeneity arising from plasmonic mode confinement, retardation, and nanoparticle-to-nanoparticle variability \cite{Peruffo2023,Yankovich2019}. 
Peruffo et al. studied the conformation of porphyrins bound to Au nanoparticles of different sizes under a variety of conditions and concluded that the orientation of the molecule can vary from perpendicular to planar with respect to the surface of the nanoparticle, binding on the metallic surface or via the surface capping ligands \cite{nicola_peruffo_engineering_2022,Peruffo2021,Peruffo2023}. 
Unlike cavity polaritons, plexcitons couple molecular excitations to a dominant plasmonic mode and to a continuum of metallic excitations, an interaction that is very sensitive to distance and orientation. Dark excitons can acquire finite plasmonic character through near-field coupling, plasmonic modes other than dipolar can hybridize with bright exciton states \cite{Rousseaux2020}, and strictly dark states can influence the dynamics by interfacial energy transfer. Disorder therefore strongly affects not only the strength of exciton–plasmon coupling but also the effective number of molecules contributing to the plexcitonic mode, the emergence of localized versus collective hybrid states, and the creation of dark states that can act as exciton traps. Plexcitonic photophysics reflect a disorder-controlled competition between exciton delocalization, localization imposed by dark states or uncoupled molecules, and quenching due to interfacial processes.  \newline
 
% Geometry of molecular aggregates
Organic molecular aggregates are an important class of emitters in plexcitonic assemblies \cite{Balci2016,Vasa2013,Deshmukh2022}. Depending on the relative angle between the transition dipole moment of its constituents, the aggregates fall into J-aggregates, H-aggregates, or X-aggregates \cite{nicholas_j_hestand_expanded_2018,Ma2021}. 
% XXX\cite{Hestand2011,Ma2021}. 
%
\begin{figure}[h!]
    \centering
    \includegraphics[width=0.5\textwidth]{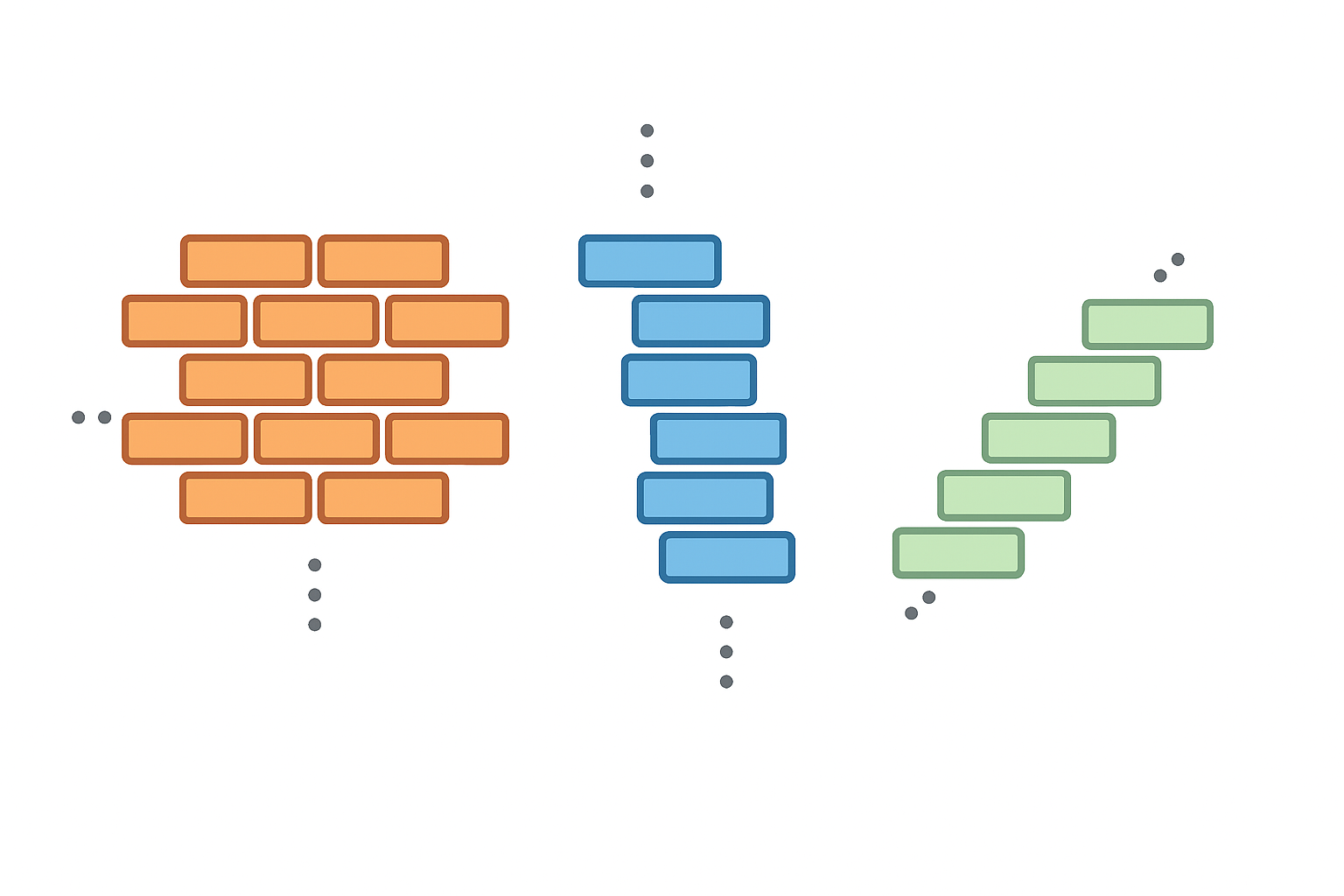}
    \caption{Possible geometries for J-aggregates are a) brickwork, b) ladder and c) staircase.}
    \label{fig:aggregation_geometries}
\end{figure}
J-aggregates concentrate the individual molecular transition dipole moment into a very bright collective transition that is red-shifted with respect to the monomer, with a very large fluorescence quantum yield and a small Stokes’ shift \cite{nicholas_j_hestand_expanded_2018}. 
% XXX,Ma2021, Anantharaman2021, Barotov2022}. 
The formation of these structures depends on the intrinsic molecular properties such as $\pi-\pi$ interactions, hydrogen bonding, and steric repulsion, as well as the solvent composition, spectator ions and overall dye concentration \cite{thermodynamic_deshmukh_2020}.
%XXX \cite{Ma2021, vonBerlepsch2000a}. 
%
J-aggregates of cyanine dyes can adopt several distinct packing geometries, each defined by how adjacent chromophores laterally slip and stack. Among the most common motifs are staircase and ladder arrangements (Fig. \ref{fig:aggregation_geometries}.c and b), in which molecules form quasi-one-dimensional slip-stacked chains that differ in the direction of slippage relative to the aggregate axis. These chain-like units often act as fundamental building blocks within larger two-dimensional sheets. A third structural motif is the brickwork arrangement (Fig. \ref{fig:aggregation_geometries}.a), a two-dimensional packing characterized by larger lateral offsets between neighboring chromophores. Together, these geometries underpin the rich polymorphism observed in J-aggregate morphologies—from narrow fibrils to extended monolayers—and they provide the structural basis for the excitonic coupling that defines the optical response of these materials \cite{Anantharaman2021,Dijkstra2016}. These structural motifs are often spectroscopically nearly degenerate in the electronic domain, making vibrational or NMR probes essential for their discrimination.
% XXX\cite{Prokhorov2012,Prokhorov2015,Prokhorov2012b,KuhnBrickwork, Ma2021,Anantharaman2021}. 

\begin{figure}
    \centering
\includegraphics[width=0.95\linewidth]{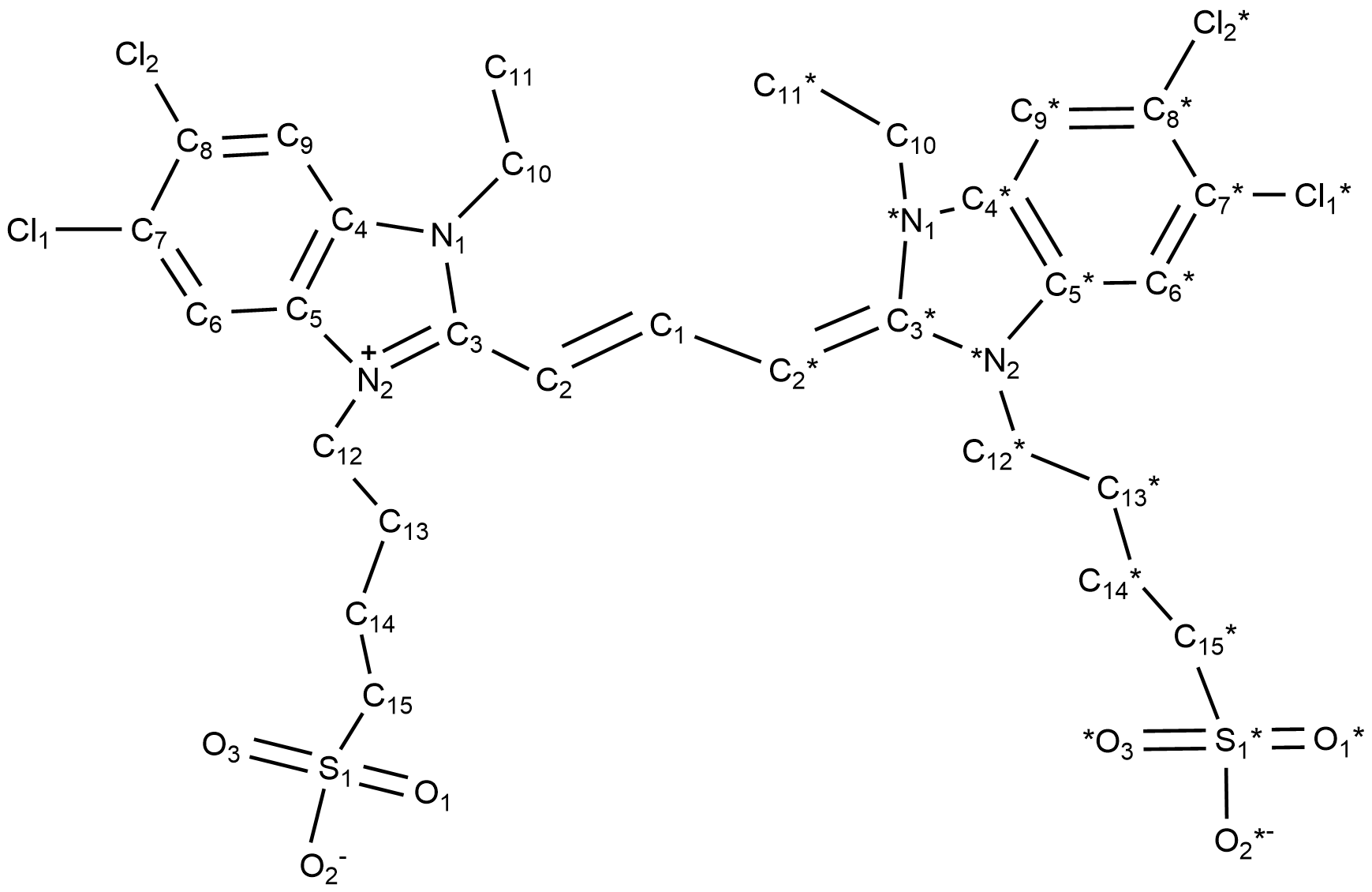}
    \caption{The chemical strucuture of TDBC. We label only non-equivalent atoms, and use the $*$ symbol to denote mirror atoms. We do not show hydrogens explicitly and will refer to them in the text according to the number of the carbon they are attached to.}
    \label{fig:ChemDraw_TDBC}
\end{figure}

% Something specific concerning TDBC
5,5',6,6'-tetrachloro-1,1'-diethyl-3,3'-di(4-sulfobutyl)-benzimidazolocarbocyanine (TDBC) is one of the workhorses of polaritonic materials due to its large oscillator strength, its well-defined J-aggregate signature,
% XXX\cite{Vasista2022} 
(Fig. \ref{fig:ChemDraw_TDBC}). It is a cyanine dye with a 3-carbon bridge between the two benzimidazole rings. Its longest aliphatic chain terminates in sulfonate groups that provide a strong attachment on metallic surfaces, which have made it a desirable molecule in plexcitonic systems as well.
Bawendi and co-workers recently synthesized a silica-capped TDBC aggregate, and on the basis of the STM-determined molecular layer height, concluded that the dye aggregated in a planar arrangement with the dye alternating the orientation of the long aliphatic chains \cite{thanippuli_arachchi_bright_2024}. 
Coles et al. carried out an extensive Raman scattering investigation of TDBC in its monomeric (preferred configuraton in methanol) and aggregated (preferred configuration in water) form \cite{coles2010characterization}. By screening a number of conformers with Density Functional Theory, they were able to suggest the most likely conformers.

In this work, we combine high-field NMR, resonance Raman and THz-Raman spectroscopy supported by density functional theory to interrogate the molecular structure of TDBC across three distinct regimes: isolated monomers in methanol, J-aggregates in aqueous solution, and Ag-bound plexcitonic assemblies. By directly comparing these reference states, we identify spectroscopic signatures that report on molecular conformation, intermolecular packing, and adsorption-induced distortions at the metal interface. In particular, NOESY cross-peaks provide constraints on side-chain symmetry and aggregate packing, while vibrational modes spanning the Raman and THz-Raman regimes selectively probe short-range molecular geometry and longer-range collective order, respectively. Together, these complementary probes allow us to constrain the most probable molecular arrangements in TDBC–Ag plexcitons and to identify experimentally accessible markers of structural heterogeneity relevant for exciton–plasmon hybridization. 

\section{Methods}

\textbf{Materials}. 5,5',6,6'-tetrachloro-1,1'-diethyl-3,3'-di(4-sulfobutyl)-benzimidazolocarbocyanine (TDBC) was purchased from Few Chemicals. For NMR experiments, the dye was purified following a reported procedure \cite{thanippuli_arachchi_bright_2024,barotov_nearunity_2022}. Briefly, TDBC is dissolved in anhydrous methanol and the solvent slowly evaporated. The supernatant is discarded and the precipitate consists of pure TDBC as seen in NMR. 
All other chemical reagents were purchased from Sigma Aldrich and used without further purification. \newline

\textbf{Aggregate preparation.} For either monomeric or J-aggregates, TDBC was dissolved in either methanol and water, respectively, and sonicated for 5 minutes prior to use. For resonance Raman measurements, a concentration of 0.5 mg/mL was used, following earlier work \cite{coles2010characterization}. The samples were then placed in a liquid nitrogen bath to avoid radiation damage. For THz-Raman measurements, 5 mg/mL samples were prepared and measured at room temperature in a confocal microscope geometry.  \newline 

\textbf{Synthesis of nanoparticles.} The synthesis of silver and plexitonic nanoparticles was carried out as reported in previous studies \cite{balci2019colloidal, balci2013ultrastrong, Balci2016}. A seed solution of Ag nanoparticles was prepared as follows. 5 mL of a 2.5 mM trisodium citrate solution and 0.25 mL of 500 mg/L poly(sodium 4-styrenesulfonate) (PSS) were initially combined with 0.3 mL of freshly made 10 mM NaBH$_4$. Subsequently, 5 mL of 0.5 mM AgNO$_3$ was gradually added to this mixture at a rate of approximately 2 mL per minute, under vigorous stirring. After 30 minutes, a yellow-colored silver nanoparticle colloid was visible.
The next step was to mix 5 mL of Millipore with 75 $\mu$L of 10 mM ascorbic acid and 60 $\mu$L of seed solution. Subsequently, 3 mL of 0.5 mM AgNO$_3$ was added dropwise ($\approx$1 mL/min) to this mixture under vigorous stirring. During addition, the colloid color transitioned from yellow to red, and finally to blue, indicating the formation of nanoprisms with a resonance wavelength near 750 nm. Finally, 0.5 mL of 25 mM trisodium citrate solution was added to stabilize the Ag nanoprisms. All reactions were performed in aqueous medium at room temperature using Millipore water throughout. The obtained nanoparticles were heated in an oil bath at 90$^{o}$C for 90 minutes. During heating, a blue shift in the nanoparticle resonance was observed, culminating in a final resonance wavelength centered at 600 nm. Nanoparticles were centrifuged for 15 min. at 7000 RPM in order to remove any remaining seed. \newline

\textbf{UV Vis.} Optical absorption was measured in an Agilent Cary 100 UV-Vis Spectrometer in a 1 mm quartz cuvette. \newline

\textbf{NMR}. ${^1}$H NMR were recorded at 298 K on a Bruker 700 MHz AVANCE III HD NMR spectrometer, equipped with a (H-C/N-D) cryoprobe. TDBC was dissolved in deuterated water at a concentration of 5 mg/mL, in deuterated methanol at a concentration of 2.5 mg/mL, and in a 6:4 (v/v) deuterated methanol–deuterated water solvent mixture at a concentration of 2.5 mg/mL. All samples were sonicated prior to measurement, and no precipitate or scattering was observed. \newline 

\textbf{Resonance Raman Spectra} were recorded at 77 K using an LN2 flow cryostat (Air Liquide, France). A diverse array of laser sources was utilized to provide excitation at various wavelengths. A Coherent Ar$^{+}$ (Sabre) laser generated 457.9, 488.0, 501.7, 514.5, and 528.7 nm. Output laser powers of 10–100 mW were attenuated to $<$ 5 mW at the sample. Scattered light was focused into a Jobin-Yvon U1000 double-grating spectrometer (1800 grooves/mm gratings) equipped with a red-sensitive, back-illuminated, LN2-cooled CCD camera. Sample stability and integrity were assessed based on the similarity between the first and last Raman spectra. Immersion in LN2 prevents thermal degradation. \newline 

\textbf{THz-Raman Spectra} were acquired at room temperature in a home-built setup. A HeNe laser at 633 nm was used to excited the sample via a wide aperture microscope. Bragg filters (Optigrate) before the sample were used to remove any parasitic radiation around the main line, and after the sample to suppress Rayleigh scattering. The scattered radiation was focused onto an Andor Kymera spectrometer equipped with a 1200 l/mm and a TE-cooled iDus CCD camera. Calibration of the THz-Raman region was done with a sulphur standard. \newline

\textbf{Computational details}. DFT calculations were carried out in Gaussian 16 \cite{g16} with the B3LYP functional \cite{Becke1993_JCP,Lee1988_PRB}
, the def2-SVP basis set \cite{Weigend2005_PCCP} with W06 density fitting \cite{Weigend2006_PCCP} and with D3(BJ) dispersion corrections \cite{Grimme2010_JCP,Grimme2011_JCC} to account for van der Waals interactions \cite{Weigend2005_PCCP,Weigend2006_PCCP}. 
The periodic surface slab calculations were performed in the Vienna Ab-initio Simulation Package (VASP) \cite{Kresse1993_PRB,Kresse1996_PRB,Kresse1996_CMS} using the RPBE functional \cite{Hammer1999_PRB}.
The solvent dielectric function was accounted for with a solvent reaction field model (SCRF) using the IEF-PCM method. The frequency of the normal modes was calculated and rescaled by 0.98 \cite{coles2010characterization}. 
The Raman activities, $S_i$ corresponding to a mode of frequency $\nu_i$,  obtained from Gaussian, were
transformed into Raman intensities, $I^{R}_{i}$, following
\begin{equation}
    I^{R}_{i}=\frac{C(\nu_i-\nu_0)^{4}S_i}{\nu_i\left[1-\exp\left(-\frac{hc\nu_i}{kT}\right)\right]}
\end{equation}
where $c$ is the light speed and $\nu_0$ is the excitation laser frequency. The constant $C$ has a value of $1\times10^{-17}.$

\section{Results}

\subsection{Synthesis and optical absorption spectroscopy}

\begin{figure}[h]
    \centering
\includegraphics[width=1.0\linewidth]{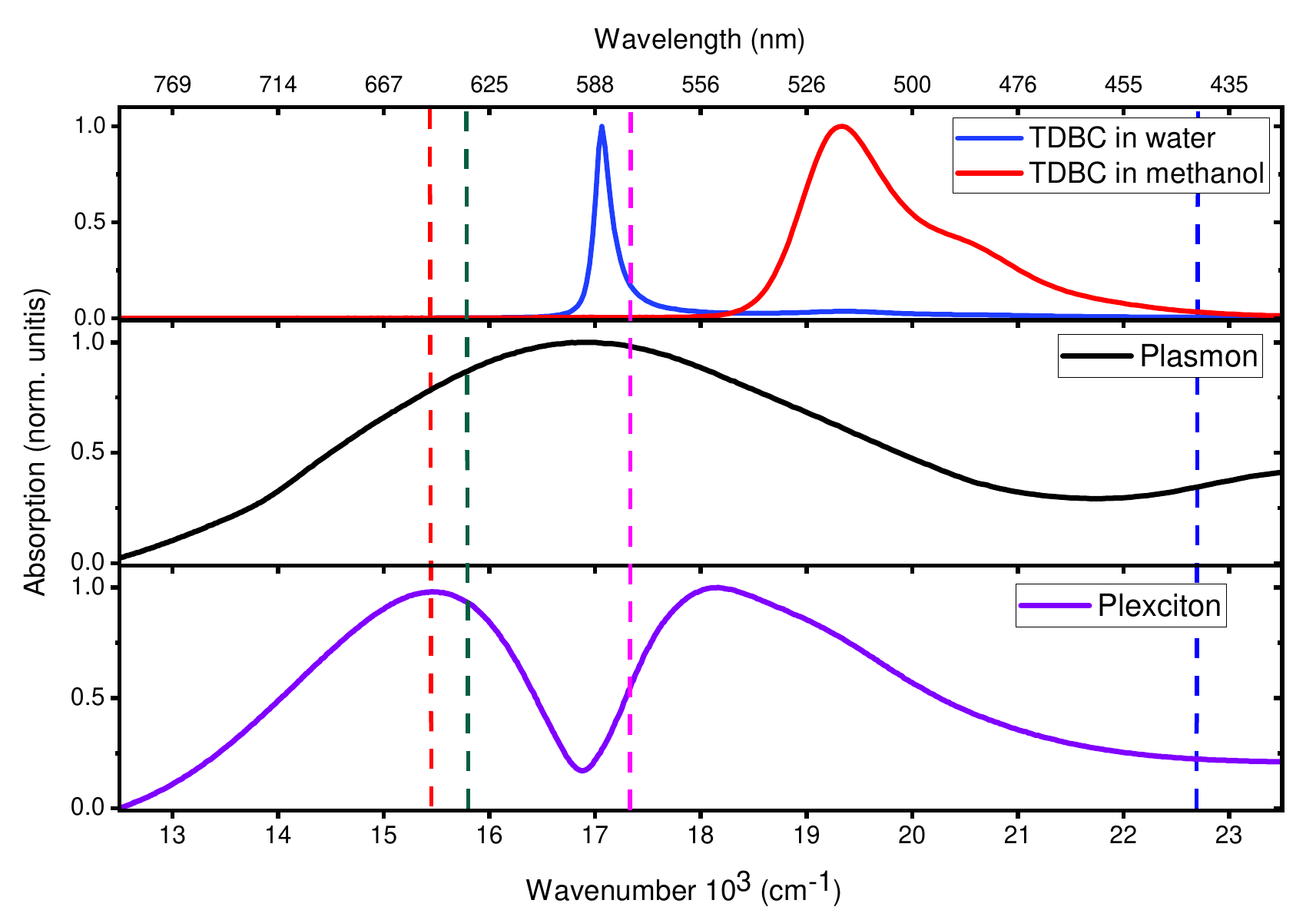}
    \caption{Normalized absorption spectra of TDBC in water (blue), TDBC in methanol (red), Ag nanoparticles (black), and the plexciton (purple). Dotted lines indicate the laser excitation wavelengths. For Raman measurements, the red, pink, and blue lines correspond to excitation wavelengths of 647, 577, and 441 nm, respectively, while the green line denotes the THz-Raman excitation wavelength at 633 nm.}
    \label{fig:UVvis}
\end{figure}

Silver nanoparticles (Ag NPs) were synthesized following previously reported procedures (see Methods and Refs.~\cite{balci2019colloidal, balci2013ultrastrong, balci2016tunable}). The resulting colloidal suspension exhibits a plasmonic absorption band centered around 600~nm (Fig.~\ref{fig:UVvis}).
Plexcitonic assemblies were prepared by adding 100~\textmu L of a 0.1~mM TDBC solution per mL of the Ag NP colloid. The mixture was centrifuged for 15~min at 7000~RPM to remove unbound dye, and the resulting precipitate was redispersed in water. Upon adsorption of TDBC onto the Ag nanoparticles, the plasmonic resonance undergoes a clear spectral splitting, consistent with the formation of hybrid exciton--plasmon states (Fig.~\ref{fig:UVvis}).
For comparison, Fig.~\ref{fig:UVvis} also shows the absorption spectra of monomeric TDBC in methanol, characterized by a vibronic progression, as well as that of aggregated TDBC in water, which displays the expected red-shifted and narrowed absorption relative to the monomer.
These four spectra define the reference electronic environments against which the vibrational and structural signatures discussed below are interpreted.

\subsection{NMR}

\begin{figure*}[tb]
    \centering
    \includegraphics[width=0.99\linewidth]{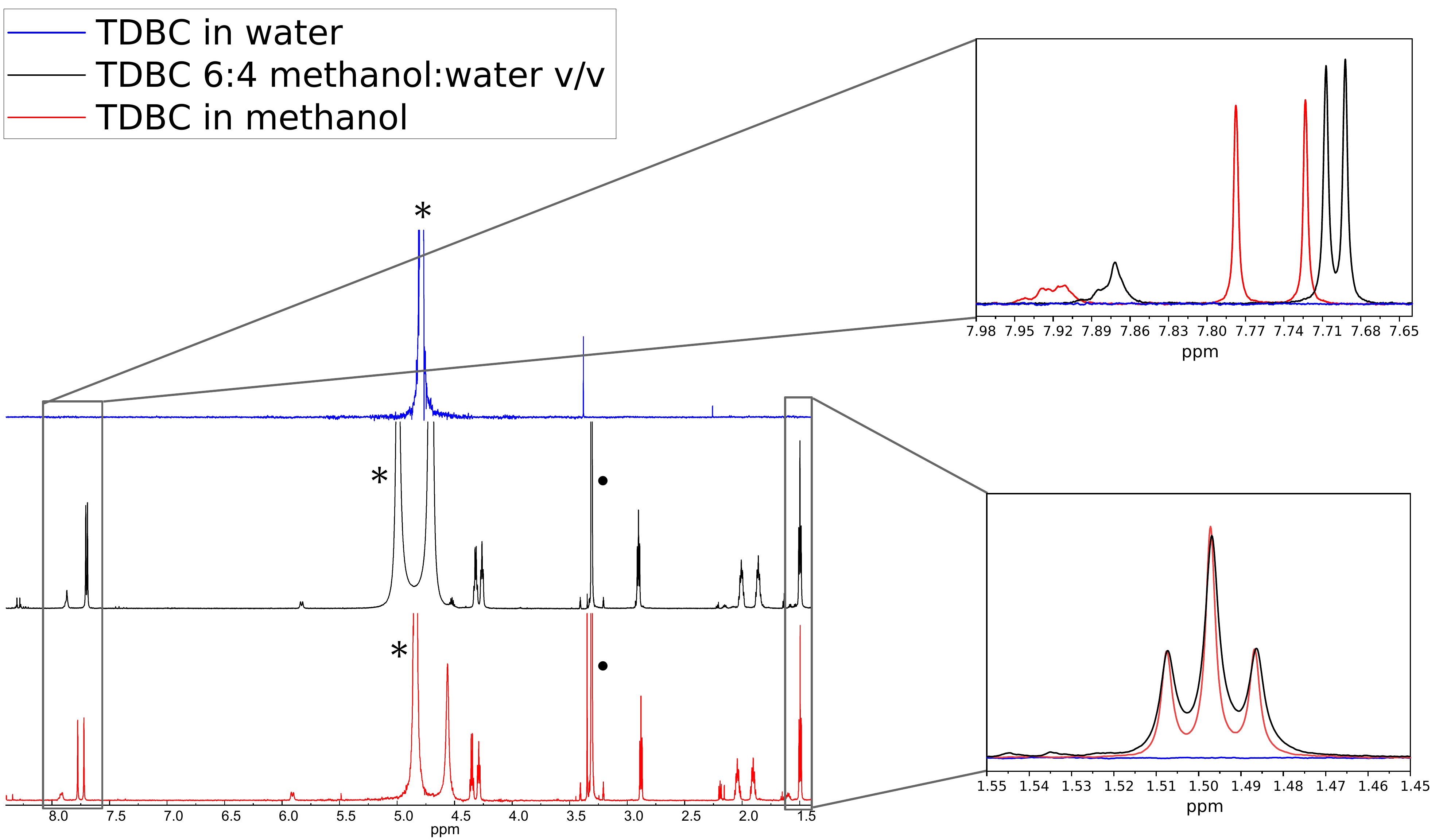}
    \caption{1H spectra of TDBC in methanol (red), in a 6:4 (v/v) MeOH–H$_2$O mixture (black) and in water (blue). The H$_2$O peak is indicated by an asterisk, while the MeOH peaks are marked with a circle. The insets show chemical shift changes and aggregation induced broadening.} 
    \label{fig:TDBC_clean}
\end{figure*}

\textbf{$^{1}$H NMR.} We recorded the $^1$H spectra of TDBC monomers, J-aggregates, and plexcitons. We found a significant difference between the spectra of TDBC as received or purified following the work of Ref. \cite{barotov_nearunity_2022}. We report the spectra of the as-received samples in the Supporting Information and those of the purified TDBC in the main text. The $^1$H resonances of the monomer in MeOH are consistent with those previously reported \cite{aviv2015synthesis} (see Figure \ref{fig:TDBC_clean} and SI). We observe a high degree of sensitivity of the areas of the H$_2$ and H$^{*}_2$ resonances depending on solvent and purity of the sample.  
%
% MAYBE ADD THIS
%The apparent integrals of the H12 and H14 resonances depend strongly on solvent composition and sample purification, and in some conditions these resonances are partially suppressed. Because no evidence of chemical degradation is observed, we attribute this behavior to changes in aggregation state and/or exchange broadening between distinct local environments, which can render a fraction of molecules NMR-invisible on the timescale of the experiment.
%
In pure water, we observe a complete disappearance of the signal, despite the solubility threshold of TDBC in water being higher than in MeOH and the absence of increased light scattering indicative of precipitation (Fig.~\ref{fig:TDBC_clean}).
We therefore attribute the signal loss to the formation of very large, water-soluble aggregates whose rotational correlation time, scaling approximately as the cube of the aggregate size, leads to homogeneous broadening beyond detection.
To restrict the aggregate size, we dissolve TDBC in a 6:4 (v/v) methanol–water mixture. We observe several chemical shift changes as well as a broadening of all peaks indicative of an increased the rotational correlation time expected for aggregate species. This is further supported by a change in the relative sign of the NOESY cross-peaks (see below) which confirms that the observed signal is representative of the aggregated species. 
$^1$H NMR spectra of plexcitonic assemblies yield a single broadened signal in the aromatic region and a congested set of resonances in the aliphatic region (see Supporting Information). Owing to the presence of stabilizing ligands on the Ag nanoparticles, the expected increase in effective aggregate size in water, and the large hydrodynamic radius of the nanoparticle–dye complex, these spectra are dominated by severe line broadening and overlap. As a result, we do not attempt a quantitative structural interpretation of the plexciton NMR data.
\newline

%%%%%%%%%%%%%%%%%%%%%%%%%%%%%%%%%%%%%%%%
%
% Dipolar coupligns for a bent geometry

\textbf{Dipolar coupling in the aliphatic–aliphatic region.} Cross-peaks in a NOESY spectrum provide information on spatial proximity between spins and provide evidence towards the determination of the preferred molecular conformations. For the monomer of TDBC in methanol, negative cross-peaks are observed, which is expected for small molecules (see SI). We observe the expected cross-peaks arising from intramolecular neighboring spins and use it as a benchmark to study aggregation. 

\begin{figure}[h!]
\centering

\subfloat[]{%
  \begin{minipage}[t]{\columnwidth}
    \centering
    \includegraphics[width=0.45\columnwidth]{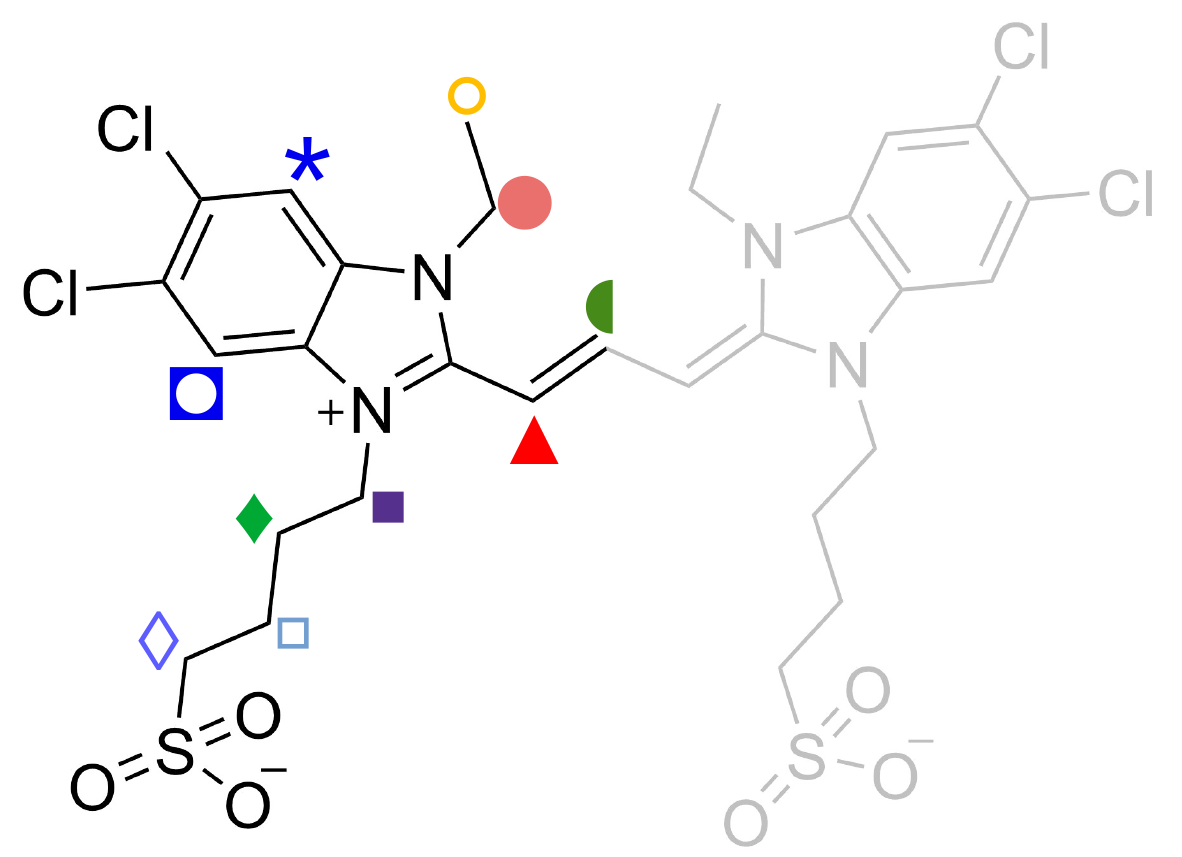}
    \label{fig:mitad_TDBC2}
  \end{minipage}%
}\\[0.5em]

\subfloat[]{%
  \begin{minipage}[t]{\columnwidth}
    \centering
    \includegraphics[width=0.85\columnwidth]{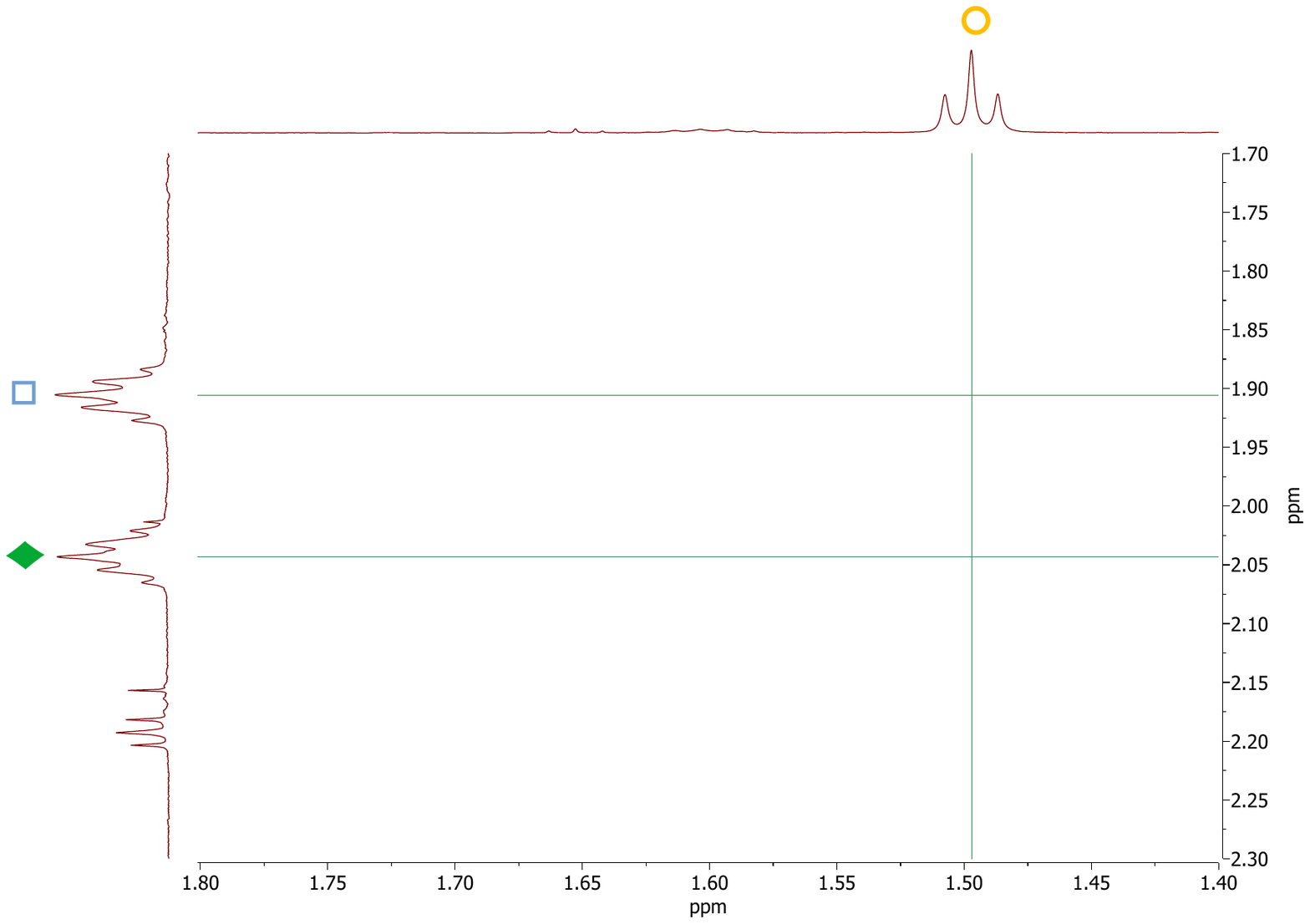}
    \label{fig:NOESY_MeOH_alif}
  \end{minipage}%
}\\[0.5em]

\subfloat[]{%
  \begin{minipage}[t]{\columnwidth}
    \centering
    \includegraphics[width=0.85\columnwidth]{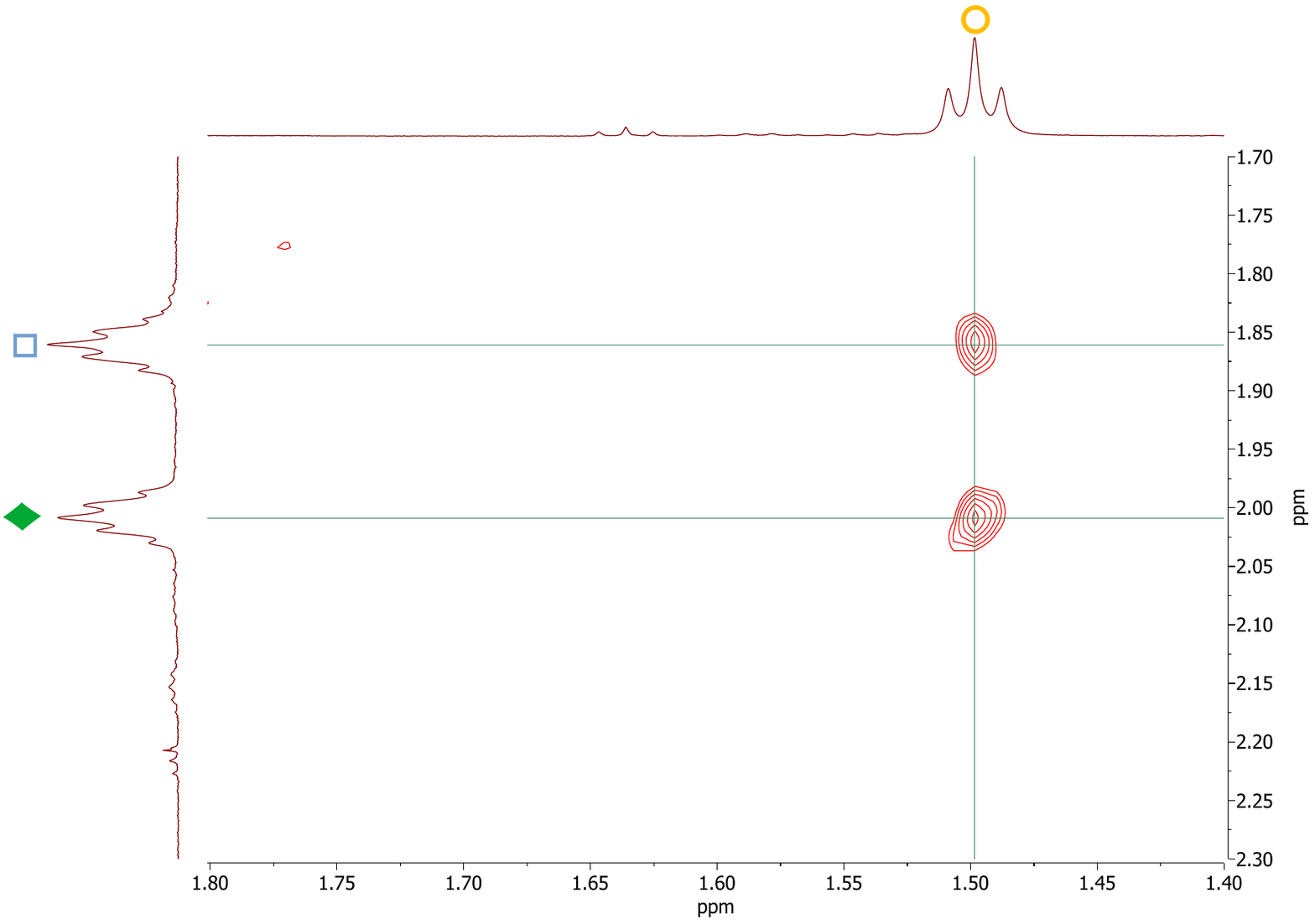}
    \label{fig:NOESY_MeOH_H2O_alif}
  \end{minipage}%
}

\caption{NOESY spectrum of TDBC in the aliphatic--aliphatic region.
(a) Molecular structure of TDBC, with one half highlighted.
(b) NOESY spectrum of TDBC in MeOH.
(c) NOESY spectrum of TDBC in 6:4 (v/v) MeOH--H$_2$O, with the appearance of cross-peaks which were not present in the monomeric TDBC.}
\label{fig:NOESYs_alif_clean}
\end{figure}

For the aggregates in a MeOH:H$_2$O mixture, we observe a sign-change in the phase of the NOESY cross-peaks which are now positive, which indicates assemblies with with much larger rotational correlation time, hence size. This confirms that the signal we are observing is coming from aggregates. We observe the appearance of new cross peaks between \(H_{11}\leftrightarrow H_{13},H_{11}\leftrightarrow H_{14}\), which are absent in the monomer (Fig.~\ref{fig:NOESYs_alif_clean}c). 
This indicates that the short and long aliphatic chains are brought into close spatial proximity within the J-aggregate, while remaining well separated in the monomer. Such correlations are consistent with an up–down–up–down packing motif for TDBC, in which the sulfobutyl chains alternate on opposite sides of the aggregation plane.
This alternating arrangement is in agreement with the structural motifs inferred from STM height measurements \cite{thanippuli_arachchi_bright_2024}.
\newline

\begin{figure}[h!]
\centering

\subfloat[]{%
  \begin{minipage}[t]{\columnwidth}
    \centering
    \includegraphics[width=0.45\columnwidth]{figures_manuscript/molecula_mitad.pdf}
    \label{fig:mitad_TDBC2}
  \end{minipage}%
}\\[0.5em]

\subfloat[]{%
  \begin{minipage}[t]{\columnwidth}
    \centering
    \includegraphics[width=0.85\columnwidth]{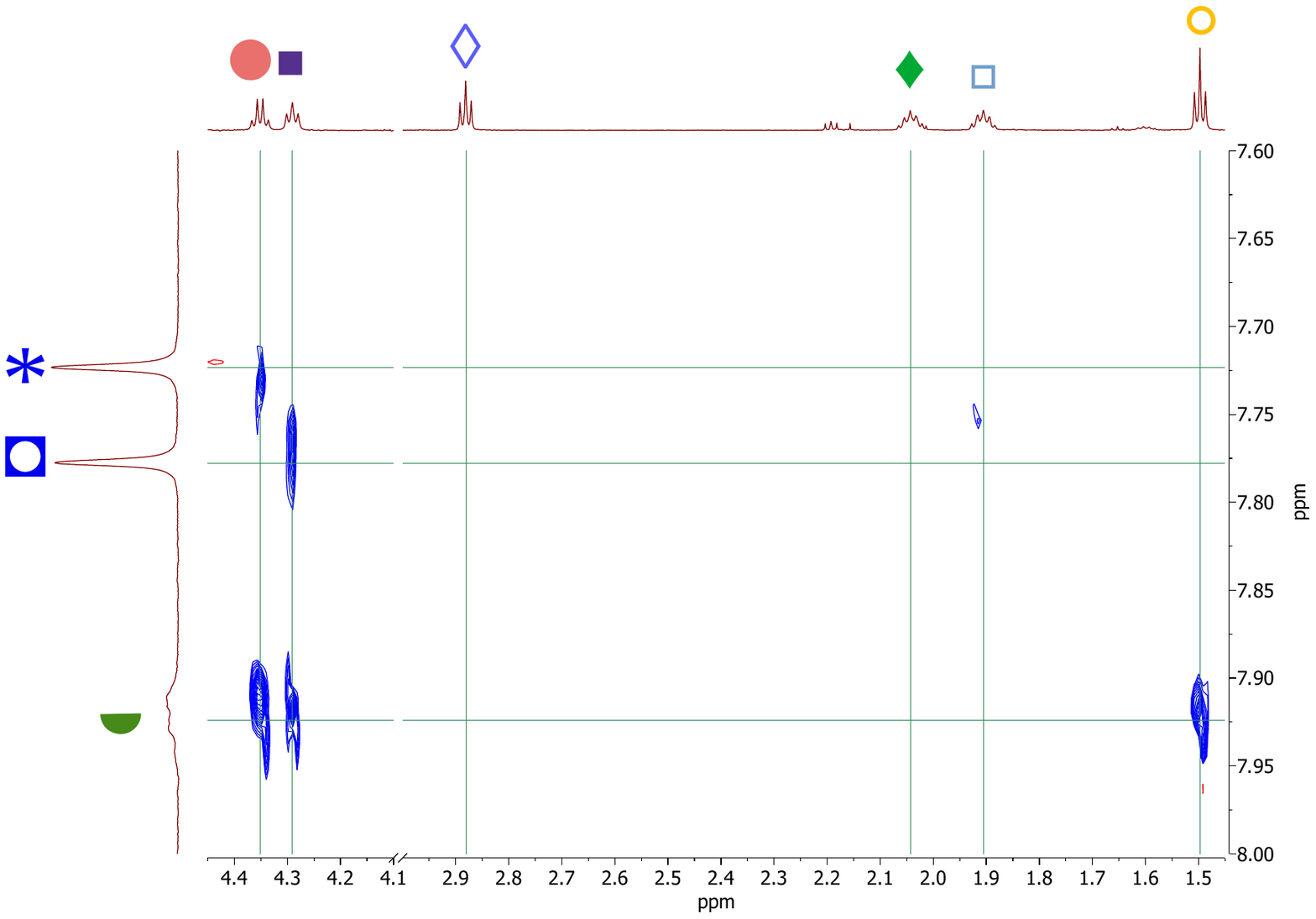}
    \label{fig:NOESY_MeOH_arom}
  \end{minipage}%
}\\[0.5em]

\subfloat[]{%
  \begin{minipage}[t]{\columnwidth}
    \centering
    \includegraphics[width=0.85\columnwidth]{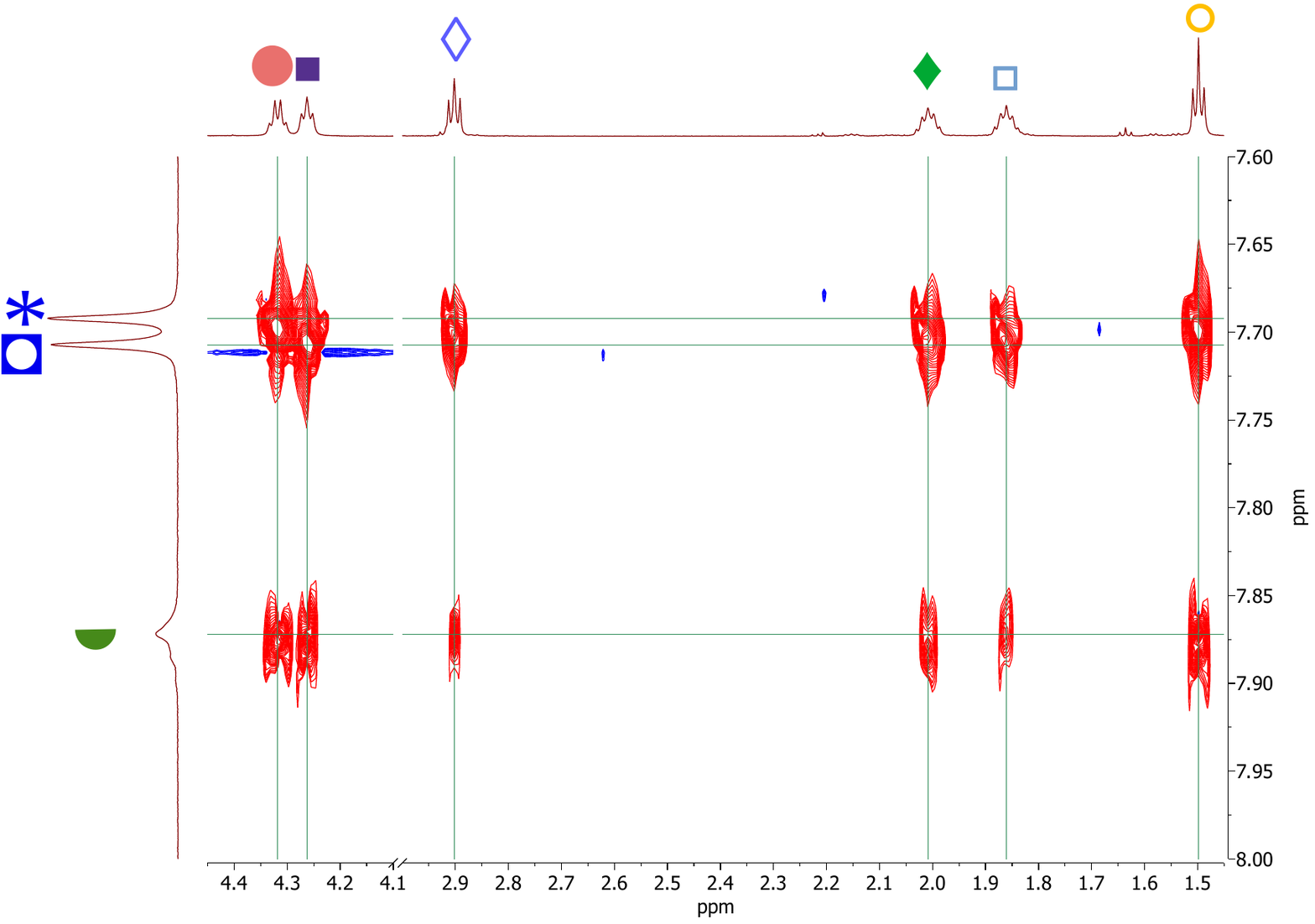}
    \label{fig:NOESY_MeOH_H2O_arom}
  \end{minipage}%
}

\caption{NOESY spectrum of TDBC in the aliphatic--aromatic region.
(a) Molecular structure of TDBC, with one half highlighted.
(b) NOESY spectrum of TDBC in methanol.
(c) NOESY spectrum of TDBC in 6:4 (v/v) MeOH--H$_2$O, with the appearance of new cross-peaks which were not present in the monomeric TDBC.}
\label{fig:NOESYs_aromatic_clean}
\end{figure}

% XXX aggregation straightens the aliphatic chain. 
\textbf{Dipolar coupling in the aromatic–aliphatic region.} When going from monomer to aggregate, we observe the appearance of several cross-peaks between \(H_{1,6,9}\) and \(H_{11,13,14,15}\) (Fig.~\ref{fig:NOESYs_aromatic_clean}). These correlations, absent in the monomer, are also consistent with an up-down-up-down arrangement although present a little more ambiguity than the aliphatic-aliphatic region due to the large number of couplings that appear. The aliphatic region provides stronger evidence towards the relative orientation of the individual monomers in the aggregate. 

Two alternative origins of the newly observed NOESY correlations are intramolecular folding and solvent-viscosity effects. Intramolecular folding is in principle possible for TDBC due to flexible alkyl chains; however, the emergence of new long-range aliphatic correlations occurs concurrently with global line broadening and an inversion of NOESY cross-peak sign, indicating entry into a slow-tumbling regime. Moreover, the viscosity change between methanol and methanol–water mixtures is modest compared to the effective increase in hydrodynamic volume expected upon aggregation. We therefore attribute the NOESY sign inversion and the new aliphatic correlations primarily to aggregation-induced packing rather than monomer folding or viscosity-driven effects.

Taken together, the $^1$H and NOESY measurements establish clear conformational and packing constraints for TDBC aggregates in solution. In particular, the emergence of aliphatic–aliphatic cross-peaks absent in the monomer, combined with the inversion of NOESY cross-peak sign, provides strong evidence for large, slowly tumbling aggregates with alternating side-chain orientations. These NMR-derived constraints will serve as a reference for interpreting the vibrational signatures of aggregation and adsorption discussed below.

\subsection{Raman} 

\subsubsection{Resonance Raman}

We acquired Raman spectra of TDBC monomers in methanol, J-aggregates in water, and plexcitonic assemblies in water, and compared them with normal modes predicted by density functional theory (DFT) calculations. The excitation wavelength of 577~nm is resonant with the aggregate monomer absorption band and overlaps the plexciton resonance.
To assist in the assignment of the observed vibrational features, we compare the experimental spectra with DFT normal-mode calculations for representative monomeric and dimeric TDBC geometries. These calculations are used here solely for vibrational assignment; the full computational methodology and adsorption geometries are discussed in the next section. Because the experimental spectra are resonance-enhanced whereas the DFT Raman activities correspond to non-resonant Raman, we primarily use the calculations to guide mode assignment and to interpret frequency shifts and mode character, while relative intensities are discussed qualitatively.
As reference monomer geometries, we consider two limiting cases: an optimized geometry with an inter-ring dihedral angle of $57^\circ$ between the planes of the two benzimidazole rings, and a planar geometry in which the two benzimidazole rings lie in the same plane. These are referred to below as the optimized and flat geometries, respectively. \newline

% Monomer
\textbf{Raman of TDBC monomer in methanol.}
We identify several Raman modes for TDBC in methanol in the 400-1700 \cm region (Table~\ref{tab:Raman_modes_monomer}). The experimental spectrum is broadly consistent with the simulated spectra of the optimized monomer (twisted benzimidazole rings with bent aliphatic chains). Two bands, near 670 and 1201~\cm, are considerably stronger than predicted. This could arise from electronic resonance enhancement not captured by non-resonant Raman activities. The 1201~\cm\ mode appears in the vibronic progression of the monomer absorption spectrum, as well as in beating frequencies observed in 2D electronic spectroscopy \cite{Finkelstein2021}, and so it is expected to have a strong resonant enhancement. 
From the simulations, it is interesting to point out that the two modes around 1598 and 1610 \cm change the order of their relative intensities depending on whether the benzimidazole rings lie on the same plane ($ I_{1610} > I_{1598} $) or with a 57$^o$ angle between them ($ I_{1610} < I_{1598} $). 
The optimized geometry, which yields a $\sim57^\circ$ dihedral angle between the aromatic ring planes, reproduces the experimental relative intensities more accurately than the planarized model. \newline

\begin{figure}[h!]
    \centering
\includegraphics[width=0.5\textwidth]{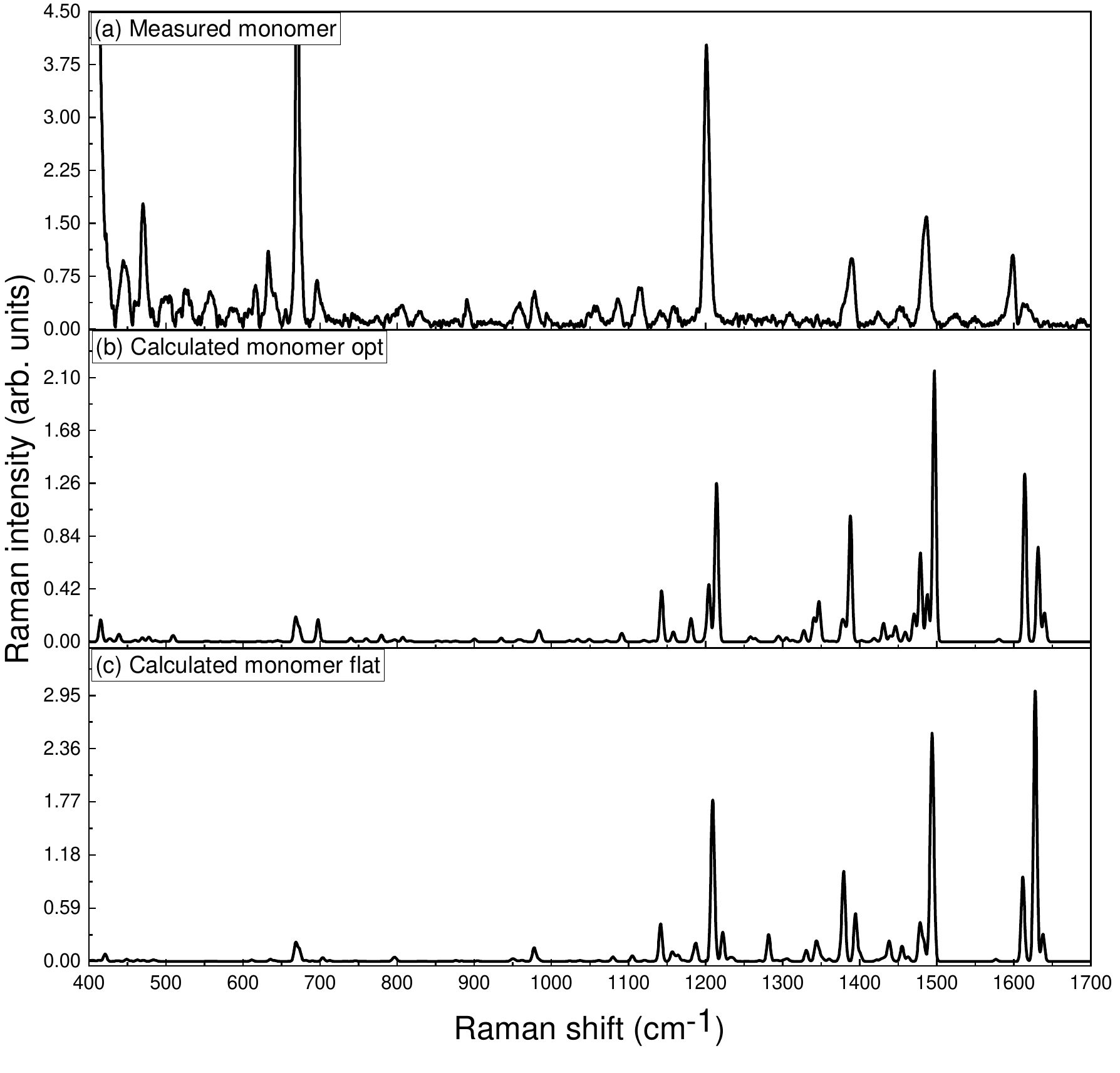}
    \caption{(a) Raman spectrum of TDBC in methanol at an excitation wavelength of 577 nm at a concentration of 0.5 mg/mL. (b) Raman spectrum of the TDBC monomer in its optimized conformation with a dihedral angle of 57$^o$. (c) Raman spectrum of the TDBC monomer with a zero dihedral angle between benzimidazole rings.}
    \label{fig:Raman_monomer}
\end{figure}

\begin{table}[!ht]
    \centering
    \begin{tabular}{|c|c|c|c|c|c|c|}
       \hline ~ &  \multicolumn{4}{c|}{calculated}  & \multicolumn{2}{c|}{experimental}\\ \hline
        ~ & \multicolumn{2}{c|}{opt}  &\multicolumn{2}{c|}{flat}  & ~ & ~ \\ \hline
        mode  & freq cm & intensity & freq cm & intensity & freq cm & intensity \\ \hline
        1 & 1632 & 0.95 & 1628 & 3.50 & 1614 & 0.42 \\ 
        2 & 1615 & 1.55 & 1612 & 1.12 & 1599 & 1.02 \\ 
        3 & x & x & x & x & 1550 & 0.18 \\ 
        4 & x & x & x & x & 1523 & 0.22 \\ 
        5 & 1497 & 2.33 & 1494 & 2.48 & 1486 & 1.57 \\ 
        6 & 1446 & 0.16 & 1455 & 0.20 & 1453 & 0.33 \\ 
        7 & 1431 & 0.17 & x & x & 1425 & 0.25 \\ 
        8 & 1388 & 1 & 1380 & 1 & 1390 & 1 \\ 
        9 & 1378 & 0.24 & x & x & 1378 & 0.33 \\ 
        10 & 1347 & 0.34 & 1347 & 0.16 & x & x \\ 
        11 & 1341 & 0.22 & 1344 & 0.23 & x & x \\ 
        12 & x & x & 1282 & 0.27 & x & x \\ 
        13 & 1204 & 0.45 & 1209 & 1.46 & 1201 & 3.79 \\ 
        14 & 1159 & 0.08 & 1157 & 0.11 & 1160 & 0.29 \\ 
        15 & 1143 & 0.34 & 1142 & 0.33 & 1144 & 0.26 \\ 
        16 & x & x & 1105 & 0.05 & 1115 & 0.61 \\ 
        17 & 1091 & 0.06 & 1081 & 0.04 & 1087& 0.43 \\ 
        18 & x & x & x & x & 994 & 0.22 \\ 
        19 & 984 & 0.07 & 978 & 0.1 & 978 & 0.52 \\ 
        20 & 960 & 0.01 & 962 & 0.01 & 959 & 0.4 \\ 
        21 & 900 & 0.01 & x & x & 891 & 0.41 \\ 
        22 & x & x & x & x & 829 & 0.27 \\ 
        23 & 807 & 0.03 & 797 & 0.03 & 805 & 0.34 \\ 
        24 & 697 & 0.09 & 703 & 0.02 & 696 & 0.66 \\ 
        25 & 668 & 0.1 & 669 & 0.1 & 671 & 6.88 \\ 
        26 & x & x & x & x & 655 & 0.28 \\ 
        27 & x & x & x & x & 642 & 0.51 \\ 
        28 & x & x & 635 & 0.01 & 633 & 1.05 \\ 
        29 & x & x & 611 & 0.01 & 617 & 0.61 \\ 
        30 & x & x & x & x & 585 & 0.34 \\ 
        31 & x & x & x & x & 557 & 0.55 \\ 
        32 & x & x & x & x & 527 & 0.63 \\ 
        33 & 509 & 0.02 & x & x & 500 & 0.51 \\ 
        34 & 469 & 0.01 & 478 & 0.01 & 470 & 1.77 \\ 
        35 & 439 & 0.02 & 449 & 0.01 & 446 & 0.88 \\ \hline
    \end{tabular}
    \caption{Frequencies and intensities of Raman modes calculated and measured for a TDBC Monomer}
    \label{tab:Raman_modes_monomer}
\end{table}

\begin{figure}[h!]
\centering

\subfloat[Raman spectrum of TDBC in water (577 nm).]{%
  \begin{minipage}[t]{\columnwidth}
    \centering
    \includegraphics[width=0.7\columnwidth]{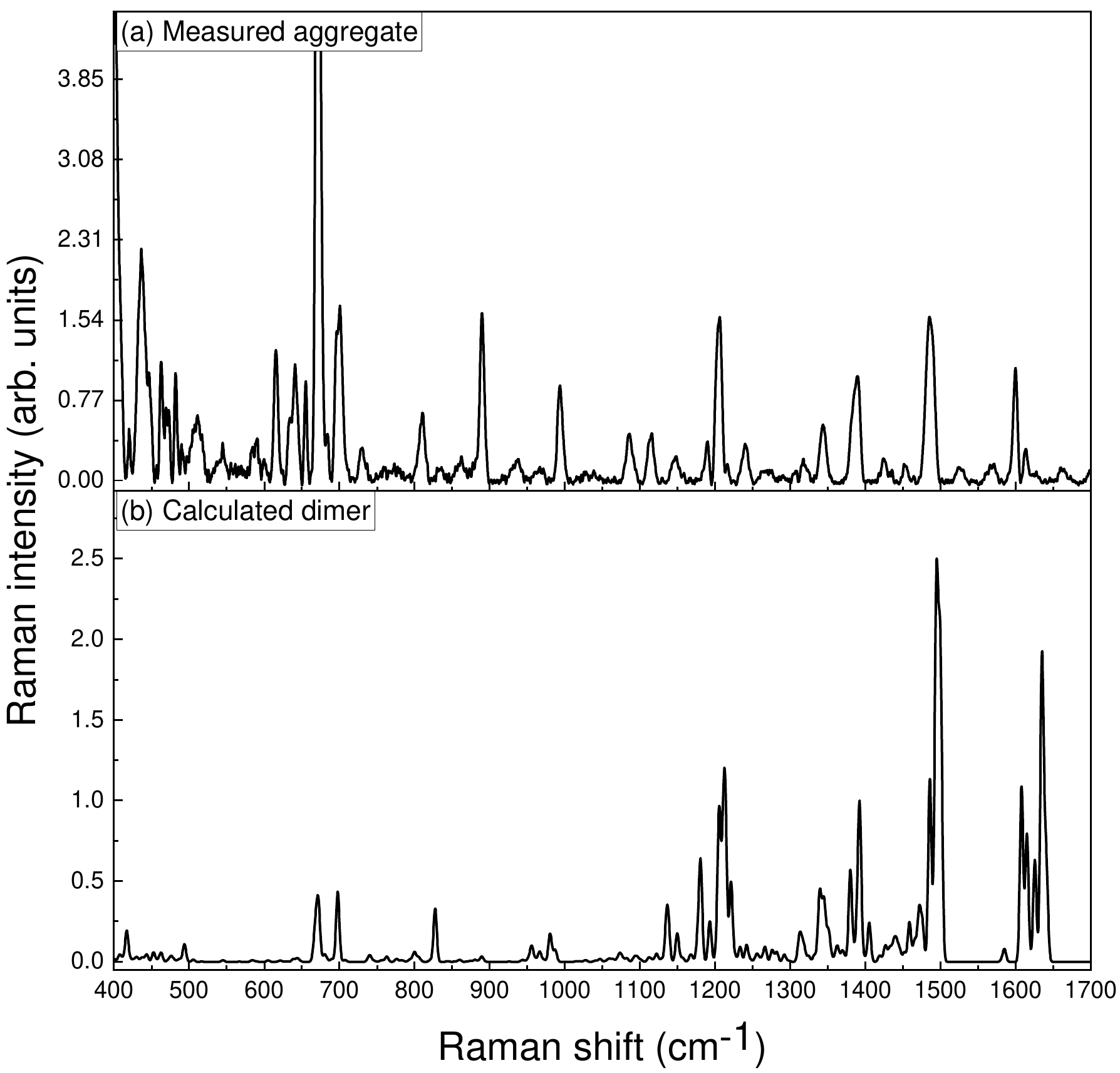}
    \label{fig:Raman_dimer}
  \end{minipage}%
}\\[0.5em]

\subfloat[800~cm$^{-1}$ region: monomer in MeOH vs J-aggregates in water.]{%
  \begin{minipage}[t]{\columnwidth}
    \centering
    \includegraphics[width=0.7\columnwidth]{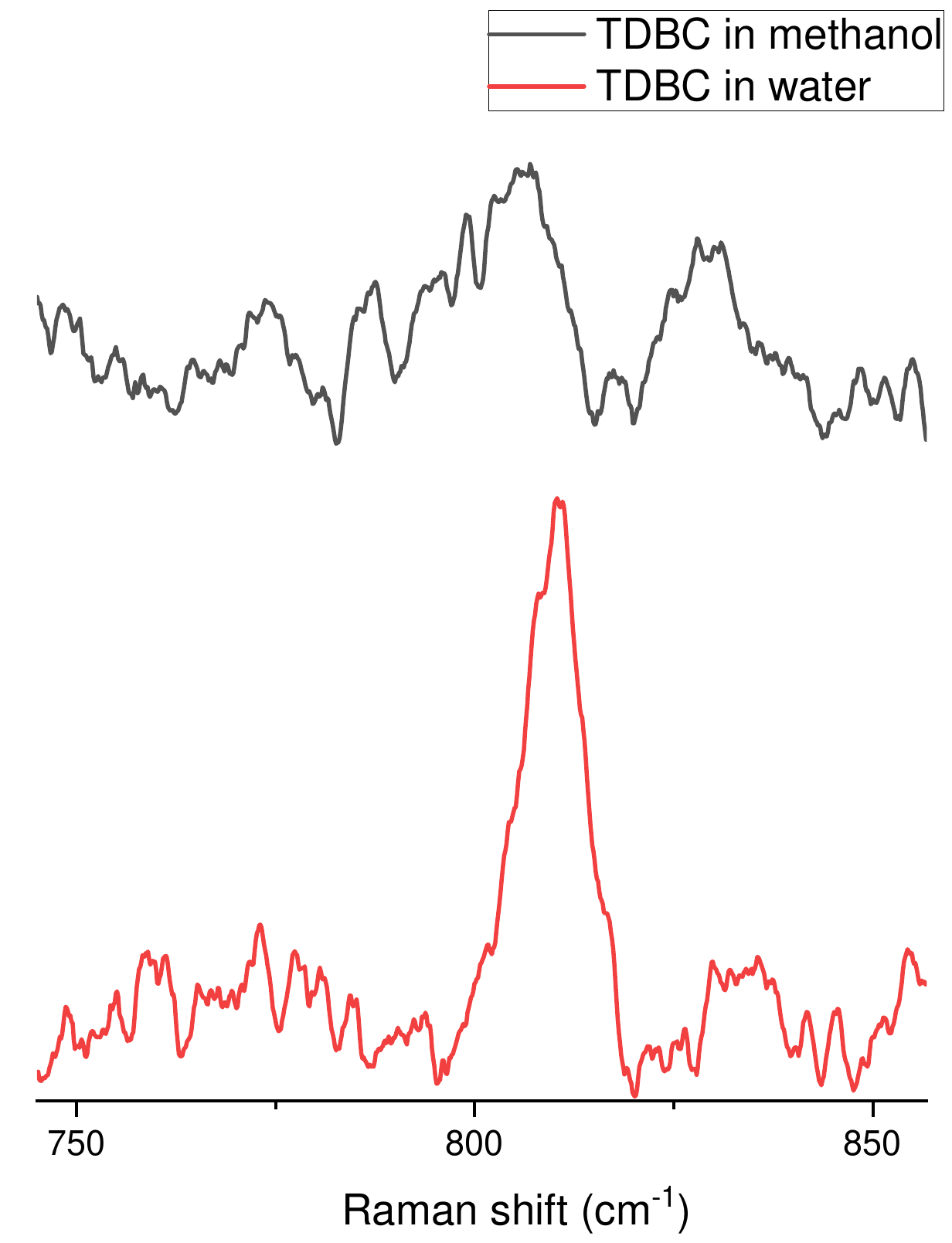}
    \label{fig:800cm_peak}
  \end{minipage}%
}

\caption{Resonance Raman signatures of TDBC aggregation in water.
(a) Raman spectrum of TDBC in water at an excitation wavelength of 577~nm (compare with dimer simulation in Fig.~\ref{fig:Raman_dimer} discussion).
(b) Zoom of the 800~cm$^{-1}$ region comparing monomeric TDBC in methanol and J-aggregates in water.}
\label{fig:Raman_water_and_800}
\end{figure}

% Water
\textbf{Raman of J-aggregates in water}.
We measured the Raman spectrum of TDBC J-aggregates in water using an excitation wavelength of 577 nm and compare with the simulated spectra obtained for a dimer structure (Fig. \ref{fig:Raman_dimer} and Table \ref{tab:Raman_H2O_agg}). 
We observe an excellent agreement between measurements and simulations in the higher frequency region. However, in the lower frequency region $<1000$ cm we observe that resonances appear much more intense than predicted, which is seen clearly for the 671 \cm mode. This could be the result of an electronic resonance enhancement, or Aggregation-Enhanced Raman Scattering (AERS) due to many more molecules participating in the aggregate \cite{enhanced_akins_2013,raman_akins_1990}. Discrepancies between experiemnts and simulations are expected given that we cannot capture the geometry of an aggregate based on a dimer structure, and simulations involving more than two interacting molecules are computationally demanding, due to both the overall size of the aggregate and the conformational flexibility of each participating molecule. As shown by Coles et al., different starting points for a dimer will give different optimized geometries and associated simulated Raman spectra \cite{coles2010characterization}.
Inspection of the atomic motions for each calculated Raman mode shows that the low frequency modes ($<$900 \cm) are delocalized over both molecules of the dimer and thus more susceptible to the specific aggregation geometry, while the higher frequency modes ($>$ 900 \cm) are localized on one of the molecules. 
% We thus can take low-frequency modes as indicators of aggregation. 
Among the different modes that change upon aggregation, and which will be mentioned in detail below, we note now that the 801 \cm region shows a double peak structure in the monomer, and becomes a single strong peak in the aggregate (see Fig. \ref{fig:800cm_peak}). The two modes at 1598 \cm and 1601 \cm are similar between monomer and aggregates. 

If the intensity ratio \(I_{1598}/I_{1612}\) is sensitive to the dihedral angle (as suggested by the flat vs optimized monomer simulations), the data are consistent with a similar average torsion in the aggregate and monomer, which would be in line with structural studies of other cyanine dyes \cite{wurthner2011j}. \newline

\begin{table}[!ht]
    \centering
    \begin{tabular}{|c|c|c|c|c|}
       \hline ~ &\multicolumn{2}{c|}{calculated} &\multicolumn{2}{c|}{experimental} \\ \hline
        ~ &\multicolumn{2}{c|}{dimer}  &\multicolumn{2}{c|}{} \\ \hline
        mode  & freq cm & intensity & freq cm & intensity \\ \hline
        1 & 1634 & 2.33 & 1613 & 0.31 \\ \hline
        2 & 1608 & 1.34 & 1599 & 0.99 \\ \hline
        3 & 1585 & 0.13 & 1568 & 0.19 \\ \hline
        4 & x & x & 1526 & 0.13 \\ \hline
        5 & 1494 & 2.87 & 1486 & 1.44 \\ \hline
        6 & 1458 & 0.34 & 1452 & 0.18 \\ \hline
        7 & 1426 & 0.16 & 1424 & 0.24 \\ \hline
        8 & 1392 & 1 & 1389 & 1 \\ \hline
        9 & 1344& 0.47 & 1344 & 0.54 \\ \hline
        10 & 1313 & 0.19 & 1317 & 0.25 \\ \hline
        11 & 1266 & 0.1 & 1266 & 0.15 \\ \hline
        12 & 1241 & 0.13 & 1240 & 0.37 \\ \hline
        13 & 1221 & 0.54 & 1216 & 0.18 \\ \hline
        14 & 1212 & 1.08 & 1205 & 1.45 \\ \hline
        15 & 1180 & 0.55 & 1189 & 0.41 \\ \hline
        16 & 1136 & 0.28 & 1147 & 0.27 \\ \hline
        17 & 1094 & 0.04 & 1114 & 0.45 \\ \hline
        18 & 1073 & 0.05 & 1086 & 0.45 \\ \hline
        19 & 980 & 0.13 & 994 & 0.85 \\ \hline
        20 & 966 & 0.06 & 966 & 0.14 \\ \hline
        21 & 955 & 0.07 & 937 & 0.24 \\ \hline
        22 & 889 & 0.03 & 890 & 1.46 \\ \hline
        23 & x & x & 862 & 0.23 \\ \hline
        24 & x & x & 835 & 0.12 \\ \hline
        25 & 827 & 0.19 & 810 & 0.61 \\ \hline
        26 & 740 & 0.03 & 730 & 0.36 \\ \hline
        27 & 698 & 0.21 & 700 & 1.52 \\ \hline
        28 & 680 & 0.05 & 683 & 0.49 \\ \hline
        29 & 671 & 0.2 & 671 & 18.15 \\ \hline
        30 & x & x & 655 & 0.96 \\ \hline
        31 & 643 & 0.02 & 641 & 1.05 \\ \hline
        32 & 638 & 0.01 & 634 & 0.65 \\ \hline
        33 & x & x & 615 & 1.13 \\ \hline
        34 & 584 & 0.01 & 589 & 0.45 \\ \hline
        35 & 545 & 0.01 & 545 & 0.4 \\ \hline
        36 & 494 & 0.04 & 511 & 0.71 \\ \hline
        37 & 476 & 0.02 & 482 & 0.98 \\ \hline
        38 & 463 & 0.02 & 463 & 1.13 \\ \hline
        39 & 443 & 0.02 & 437 & 2.07 \\ \hline
        40 & 417 & 0.06 & 420 & 0.66 \\ \hline
        41 & 407 & 0.02 & 401 & 4.17 \\ \hline
        42 & 389 & 0.02 & 387 & 0.65 \\ \hline
        43 & 374 & 0.01 & 373 & 1.42 \\ \hline
        44 & 361 & 0.03 & 355 & 1.83 \\ \hline
        45 & 331 & 0.02 & 339 & 0.58 \\ \hline
        46 & 319 & 0.02 & 324 & 5.05 \\ \hline
        47 & 315 & 0.02 & 317 & 1.97 \\ \hline
        48 & 275 & 0.03 & 274 & 0.68 \\ \hline
        49 & 221 & 0.01 & 226 & 1.43 \\ \hline
    \end{tabular}
    \caption{Raman modes for TDBC J-aggregates in water and the Raman simulated modes for a TDBC dimer.}
    \label{tab:Raman_H2O_agg}
\end{table}

\begin{figure*}[t]
    \centering
\includegraphics[width=1.0\textwidth]{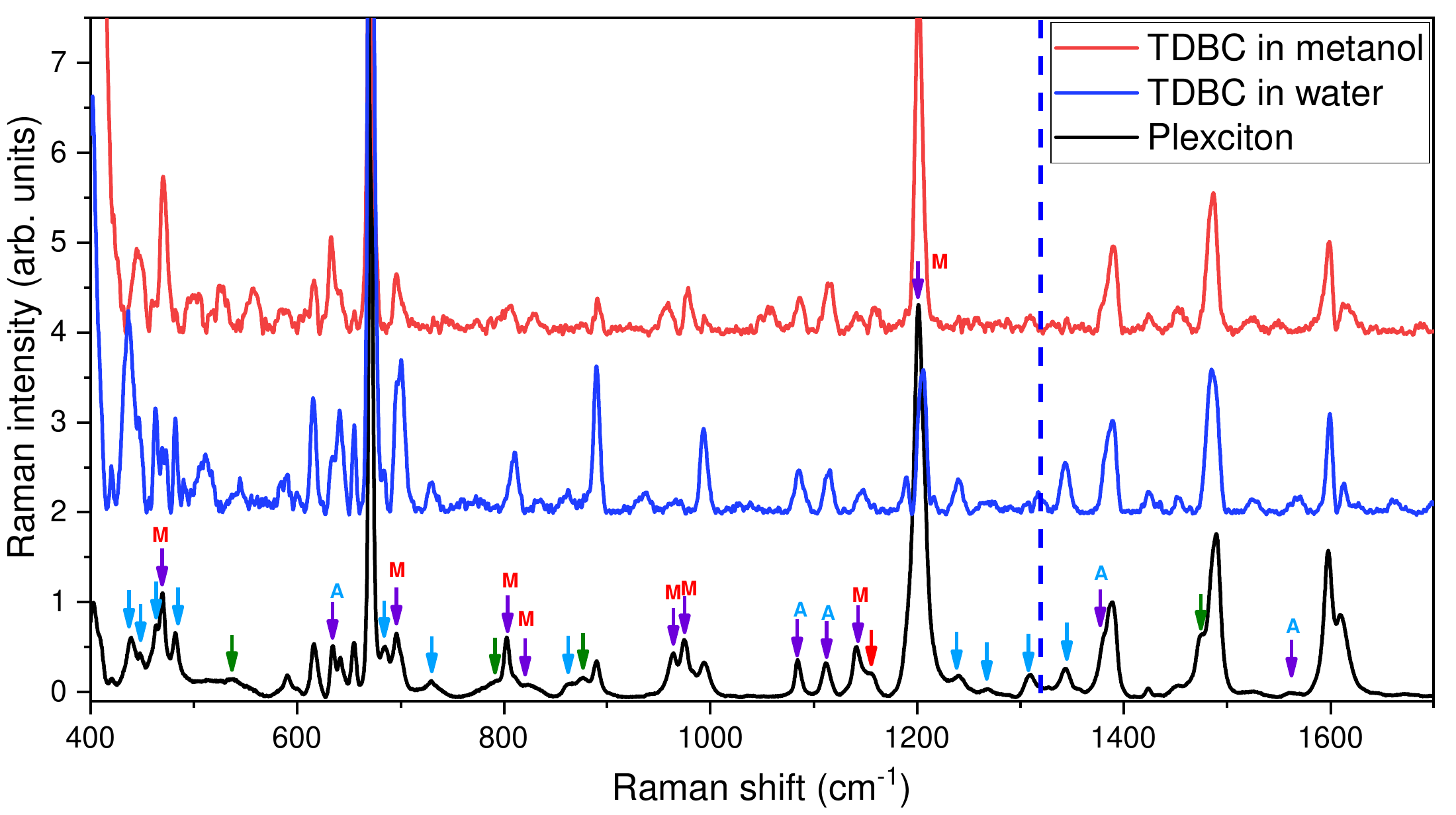}
    \caption{Raman spectra of TDBC in methanol (red), TDBC in water (blue) and TDBC-Ag plexcitonic assemblies (black). The modes that change when binding to the nanoparticle are indicated by arrows. The red arrows indicate the modes that are present in the monomeric form and plexcitonic assemblies, the blue arrows indicate the modes that are present in the J-aggregate form and plexcitonic assemblies, green arrows indicate new modes that are only seen in plexcitonic assemblies and the purple arrows indicate indicate modes that are present in all three species. The letters M and A denote a similarity of the mode frequency with monomer or aggregate, respectively. }
    \label{fig:Raman_peaks}
\end{figure*}

\textbf{Raman of plexcitons in water}. A comparison of the Raman spectra of plexcitons with that of the monomer and the aggregate shows shared features with each. Fig. \ref{fig:Raman_peaks} shows the Raman spectra of all three species. We have marked plexciton mode with a red arrow if the mode exists in the monomer, a blue arrow if it exists in the aggregate, and a purple arrow if it is present in both. For the case where the mode is present both in the monomer and the dimer, we have added an \textbf{M} to indicate that its frequency is closest to the monomer, or an \textbf{A} to indicate that it is closest to the aggregate, or left it blank if the mode's frequency is unchanged in monomer and aggregate. We have marked a green arrow if the mode appears exclusively for the plexcitons. Modes that remain invariant for TDBC in monomeric or aggregated form, and which also appear in the plexciton, are unmarked.
%
%We see 11 modes that are exclusive to water and also appear in the plexcitons, and one exclusive to the monomer that also appears in the plexciton. Some shared modes between monomer, aggregate and plexciton appear at frequencies closer to monomer (7 modes) and aggregate (5 modes), while some modes are exclusive to the plexciton (4 modes). There is one mode that appears in the monomer but not in the aggregates, and one mode that appears in the aggregates but not in the monomer. 
%
Overall, the plexciton spectrum shares many modes with the aqueous aggregate, but also retains monomer-like signatures and exhibits a small number of additional features, indicating adsorption-induced structural heterogeneity at the interface.
%
%It is worth noting that while we do see many similarities in this 400-1700 \cm between plexcitons and J-aggregates, we also see some modes which are shared between the monomer the plexcitons, as evidence that adsorption modifies the cyanine's geometry. 

\begin{figure*}[t]
    \centering
    \includegraphics[width=1.0\textwidth]{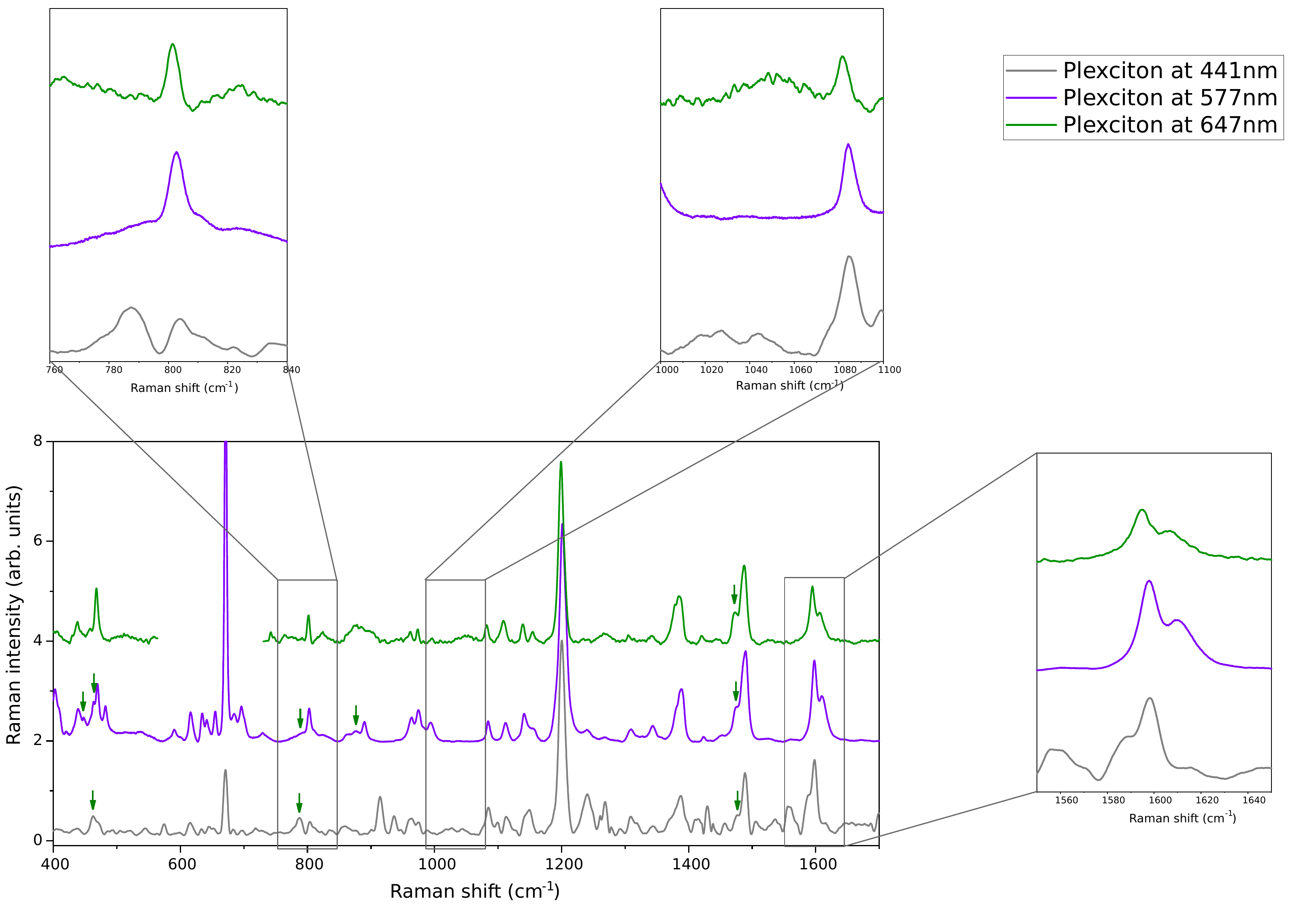}
    \caption{Raman spectra of plexciton assemblies at different excitation wavelengths. The break for the 647 nm excitation wavelength is due to a parasitic satellite line of the exciting laser.}
\label{fig:plexciton_different_wavelengths_smooth}
\end{figure*}

\textbf{Excitation wavelength dependence}. The excitation wavelength dependence of the plexciton Raman modes is shown in Fig. \ref{fig:plexciton_different_wavelengths_smooth}. The three excitation wavelengths are resonant with subsets of molecules with different electronic transition energies (for example due to different conformations of the molecules), so that we effectively select different sub-ensembles of molecules with different geometries. Additionally, exciting a plasmon with a different wavelength creates a different field configuration, enhancing the field in spatially different locations. Different Raman excitation wavelengths sample both different spatial regions, as well as molecules with different electronic excitation transitions. It provides an indication of the heterogeneity of conformations. 
We observe that the spectra at 577 and 647 nm share many similarities, with some variations in the peak intensities which is expected. 577 and 647 nm are resonant with the upper and lower polariton branches, respectively, with the wavelength at 577 nm also being resonant with the J-aggregates in solution. The Raman spectra with a 441 nm excitation wavelength shows more marked differences, as the appearance of new modes and frequency shifts of the existing ones. This excitation wavelength is resonant with the blue tail of the monomer electronic transition as well as the blue tail of the upper polariton branch so that we are sensitive to monomer and plexciton species (should they be present).   
We see indications of monomer species present when exciting with 441 nm. The region around 800 \cm displays a double peak structure, which we found was a signature of the monomer, albeit red-shifted, while for excitation wavelengths of 577 and 647 nm, we observe a single mode characteristic of the J-aggregate. Interestingly, the ratio of intensities $I_{1598}/I_{1612}$ is reverted at 441 nm than the rest of the wavelengths investigated, suggesting a planarization of the aromatic ring. 

These observations provide evidence for a minority population of monomer-like TDBC species in the plexciton sample, likely adsorbed at the Ag surface in a more planarized geometry. \newline

% THz-Raman

\subsubsection{THz-Raman}

The THz-Raman region probes low-frequency intermolecular and collective modes that are particularly sensitive to longer-range order and mesoscale packing.

\begin{figure}
    \centering
    \includegraphics[width=0.95\linewidth]{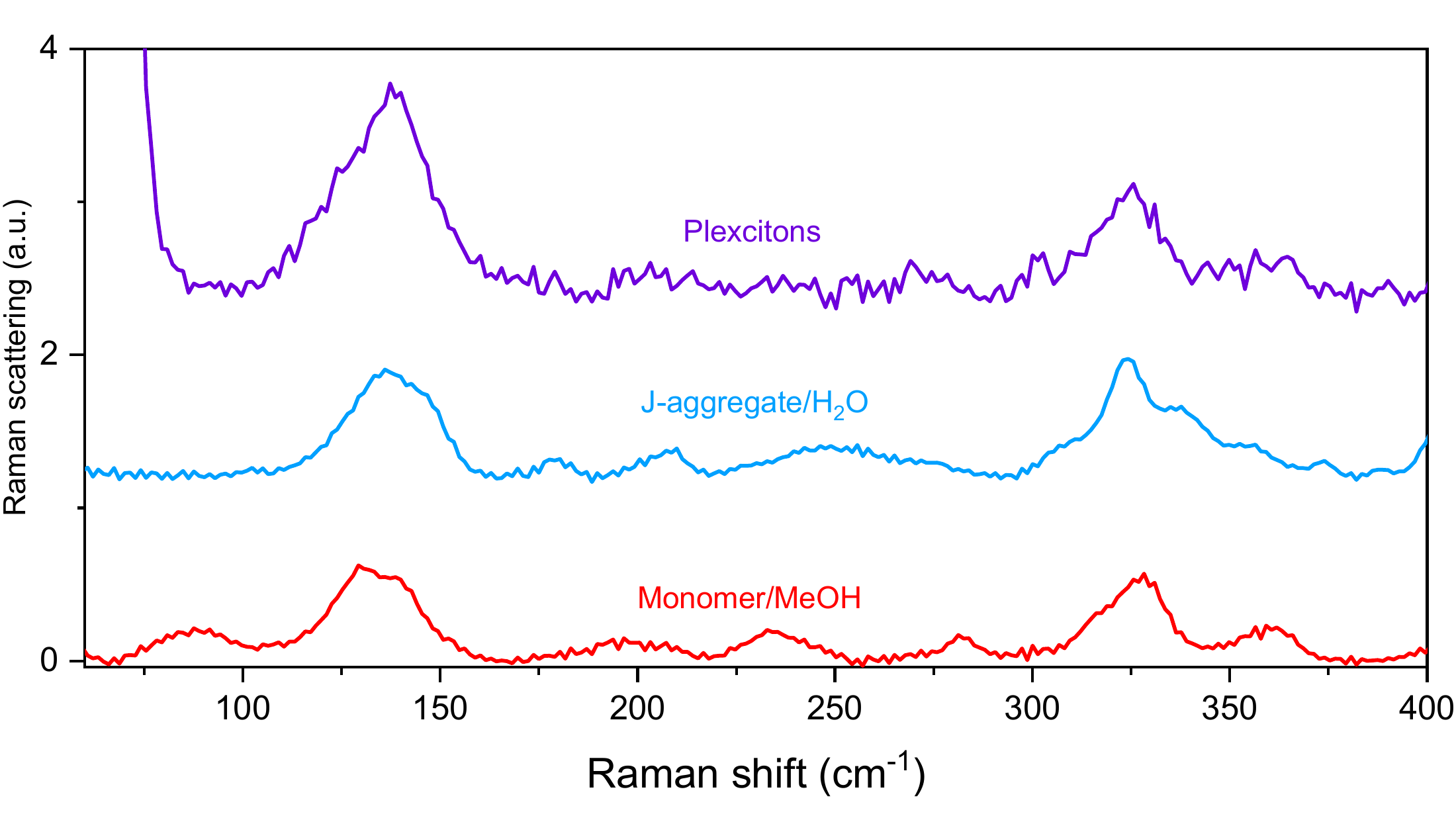}
    \caption{THz-Raman spectra of TDBC monomer in methanol (red), TDBC J-aggregates in water (blue) and  plexcitons (purple)}
    \label{fig:THz_Raman}
\end{figure}

We have also investigated the bare Ag nanoparticles, monomer, J-aggregates, and plexcitons in solution with THz-Raman spectroscopy with an excitation wavelength of 633 nm in order to probe the extended structure (see Fig. \ref{fig:THz_Raman}). We have normalized this region to the 600 \cm resonance ratios measured at 577 nm excitation, so that although the wavelength used is different, the relative intensities shown are not arbitrary and do not depend on sample preparation and concentration.
We see in this region several resonances, with two dominant features around 140 \cm and 322 \cm. The 140 \cm resonance is similar in monomers and aggregates, and becomes broader and with a different lineshape in plexcitons. However, the 322 \cm feature in plexcitons is more similar to the monomer, whereas in the aggregate we can resolve its composition in many different resonances, suggesting a longer range order than is present in the monomer or plexcitons. 
We conclude that the modes influenced by nearest neighbor interactions - probed by higher-frequencies Raman modes - remain very similar in plexcitons to that of the J-aggregate in solution, partly explaining the similarity in electronic transition frequencies of the bright state, however the long range order that the TDBC molecules adopt - probed by lower frequency modes - is distorted from those of J-aggregates in solution. 

\begin{table}[ht!]
    \centering
    \begin{tabular}{|c|c|c|c|c|c|}
    \hline
        \multicolumn{2}{|c|}{methanol}  &\multicolumn{2}{c|}{ water}&\multicolumn{2}{c|}{plexciton}  \\ \hline
        freq (\cm) & intensity & freq (\cm) & intensity & freq (\cm) & intensity \\ \hline
        1614 & 0.42 & 1613 & 0.33 & 1610 & 0.86 \\ \hline
        1599 & 1.02 & 1599 & 1.04 & 1598 & 1.58 \\ \hline
        1550 & 0.18 & 1568 & 0.2 & 1561 & --- \\ \hline
        1523 & 0.22 & 1526 & 0.13 & 1525 & 0.01 \\ \hline
        1486 & 1.57 & 1486 & 1.51 & 1489 & 1.77 \\ \hline
        x & x & x & x & 1475 & 0.65 \\ \hline
        1453 & 0.33 & 1452 & 0.17 & 1453 & 0.08 \\ \hline
        1425 & 0.25 & 1424 & 0.22 & 1424 & 0.04 \\ \hline
        1390 & 1 & 1389 & 1 & 1388 & 1 \\ \hline
        1378 & 0.33 & 1381 & 0.59 & 1381 & 0.66 \\ \hline
        x & x & 1344 & 0.57 & 1343 & 0.27 \\ \hline
        x & x & 1317 & 0.25 & 1309 & 0.19 \\ \hline
        x & x & 1266 & 0.15 & 1268 & 0.04 \\ \hline
        x & x & 1240 & 0.37 & 1240 & 0.18 \\ \hline
        x & x & 1216 & 0.18 & x & x \\ \hline
        1201 & 3.79 & 1205 & 1.5 & 1202 & 4.31 \\ \hline
        x & x & 1189 & 0.42 & x & x \\ \hline
        1160 & 0.29 & x & x & 1155 & 0.21 \\ \hline
        1144 & 0.26 & 1147 & 0.25 & 1141 & 0.5 \\ \hline
        1115 & 0.61 & 1114 & 0.47 & 1112 & 0.32 \\ \hline
        1087 & 0.43 & 1086 & 0.46 & 1084 & 0.36 \\ \hline
        994 & 0.22 & 994 & 0.9 & 994 & 0.33 \\ \hline
        978 & 0.52 & 966 & 0.16 & 975 & 0.58 \\ \hline
        959 & 0.4 & 937 & 0.25 & 964 & 0.43 \\ \hline
        891 & 0.41 & 890 & 1.53 & 890 & 0.35 \\ \hline
        x & x & x & x & 876 & 0.16 \\ \hline
        x & x  & 862 & 0.23 & 863 & 0.1 \\ \hline
        829 & 0.27 & 835 & 0.12 & 823 & 0.09 \\ \hline
        x & x & x & x & 809 & 0.19 \\ \hline
        805& 0.34 & 810 & 0.65 & 803 & 0.61 \\ \hline
        x & x & x & x & 793 & 0.13 \\ \hline
        x & x & 730 & 0.37 & 730 & 0.13 \\ \hline
        696 & 0.66 & 700 & 1.61 & 696 & 0.65 \\ \hline
        x & x & 683 & 0.49 & 685 & 0.51 \\ \hline
        671 & 6.88 & 671 & 19.02 & 671 & 7.95 \\ \hline
        655 & 0.28 & 655 & 0.96 & 655 & 0.55 \\ \hline
        642 & 0.51 & 641 & 1.09 & 642 & 0.39 \\ \hline
        633 & 1.05 & 634 & 0.68 & 634 & 0.51 \\ \hline
        617 & 0.61 & 615 & 1.21 & 616 & 0.53 \\ \hline
        585 & 0.34 & 589 & 0.52 & 590 & 0.19 \\ \hline
        557 & 0.55 & x & x & x & x \\ \hline
        x & x & 545 & 0.41 & x & x \\ \hline
        527 & 0.63 & x & x & 533 & 0.15 \\ \hline
        x & x & 511 & 0.71 & 514 & 0.14 \\ \hline
        500 & 0.51 & x & x & x & x \\ \hline
        x & x & 482 & 1.06 & 482 & 0.66 \\ \hline
        470 & 1.77 & 463 & 1.13 & 470 & 1.1 \\ \hline
        x & x & x & x & 463 & 0.75 \\ \hline
        x & x & x & x & 448 & 0.44 \\ \hline
        446 & 0.88 & 437 & 2.11 & 439 & 0.6 \\ \hline
        -- & -- & 420 & 0.66 & 420 & 0.15 \\ \hline
        -- & -- & x & x & 408 & 0.65 \\ \hline
        -- & -- & 401 & 4.36 & 403 & 0.99 \\ \hline
        -- & -- & 387 & 0.65 & 389 & 0.32 \\ \hline
        -- & -- & 373 & 1.48 & 373 & 0.68 \\ \hline
        -- & -- & x & x & 363 & 0.91 \\ \hline
        -- & -- & 355 & 1.85 & x & x \\ \hline
        -- & -- & 339 & 0.58 & 340& 1.26 \\ \hline
        -- & -- & 324 & 5.26 & 325& 3.79 \\ \hline
        -- & -- & 317 & 2.14 & -- & -- \\ \hline
        -- & -- & 274 & 0.66 & -- & -- \\ \hline
        -- & -- & 226 & 1.4 & -- & -- \\ \hline
    \end{tabular}
    \caption{Vibrational modes of TDBC in methanol, water and plexcitons.}
    \label{tab:vib_modes}
\end{table}

\subsection{DFT}

% Small intro paragraph

To rationalize the experimental NMR and Raman observations, we investigated several representative geometries of TDBC using density functional theory, including monomer conformers with different inter-ring dihedral angle between the planes of the two benzimidazole aromatic systems, dimer configurations as minimal models of aggregation, and surface-bound conformations on Ag. These calculations are used to identify energetically accessible geometries consistent with the spectroscopic constraints and to analyze the vibrational character of the associated normal modes.

\begin{figure}[h!]
     \centering
  \includegraphics[width=0.3\textwidth]{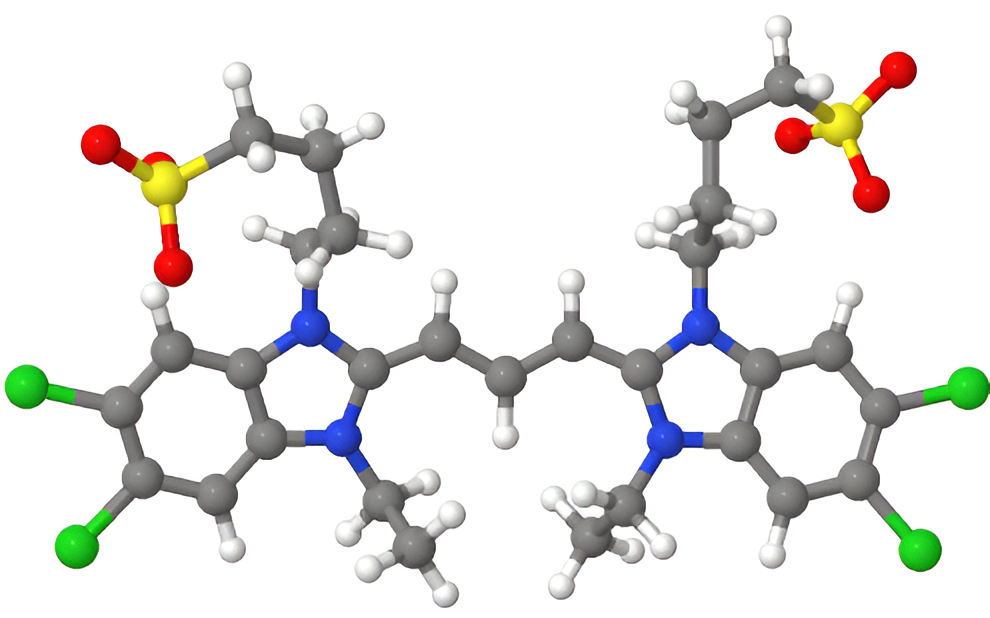}
        \caption{Geometry of the TDBC monomer structure in planar conformation. }
        \label{fig:flat}
\end{figure}

\begin{figure}[h!]
\centering

\subfloat[]{%
  \begin{minipage}[t]{\columnwidth}
    \centering
    \includegraphics[width=0.3\columnwidth]{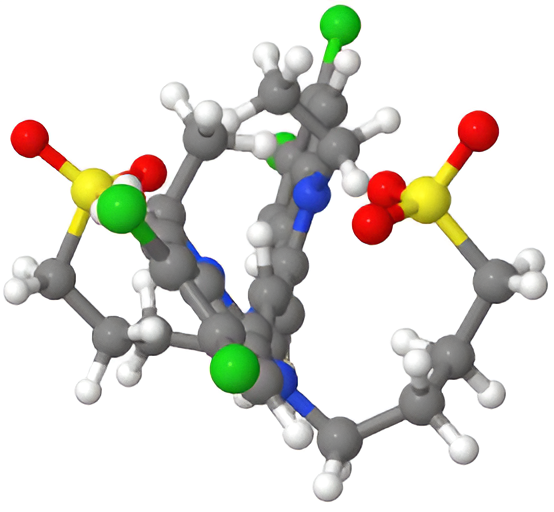}
    \label{fig:mono_opt_side}
  \end{minipage}%
}\\[0.5em]

\subfloat[]{%
  \begin{minipage}[t]{\columnwidth}
    \centering
    \includegraphics[width=0.45\columnwidth]{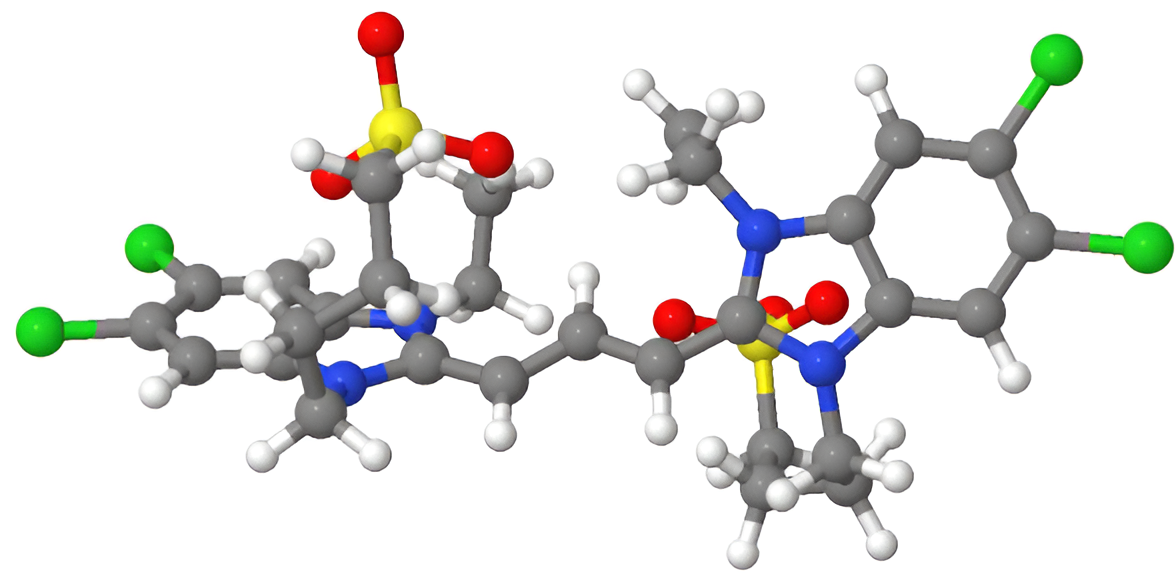}
    \label{fig:mono_opt_front}
  \end{minipage}%
}

\caption{Geometry of the TDBC monomer structure in optimized conformation.
(a) side view; (b) front view.}
\label{fig:mono_opt}
\end{figure}

% Dihedral scan
Due to the flexibility of the monomer imparted by the aliphatic chains and the torsional degree of freedom between the two rings, we carried out a geometry optimization and a scan of the dihedral angle. 
% Monomer
We selected a geometry in which the $\pi$ system is constrained to a planar conformation (flat monomer, Fig.~\ref{fig:flat}) as well as a fully optimized geometry (optimized monomer, Fig.~\ref{fig:mono_opt}) for further analysis, motivated by the sensitivity of specific Raman modes to aromatic torsion.
The two geometries differ in energy by 0.251~eV, favoring the twisted conformation.
The obtained optimized geometry agrees with that reported in \cite{coles2010characterization}. We calculate an angle between the planes of the two aromatic rings of 57$^o$. % era 57.35, lo redondee 

\begin{figure}[h!]
\centering

\subfloat[]{%
  \begin{minipage}[t]{\columnwidth}
    \centering
    \includegraphics[width=0.45\columnwidth]{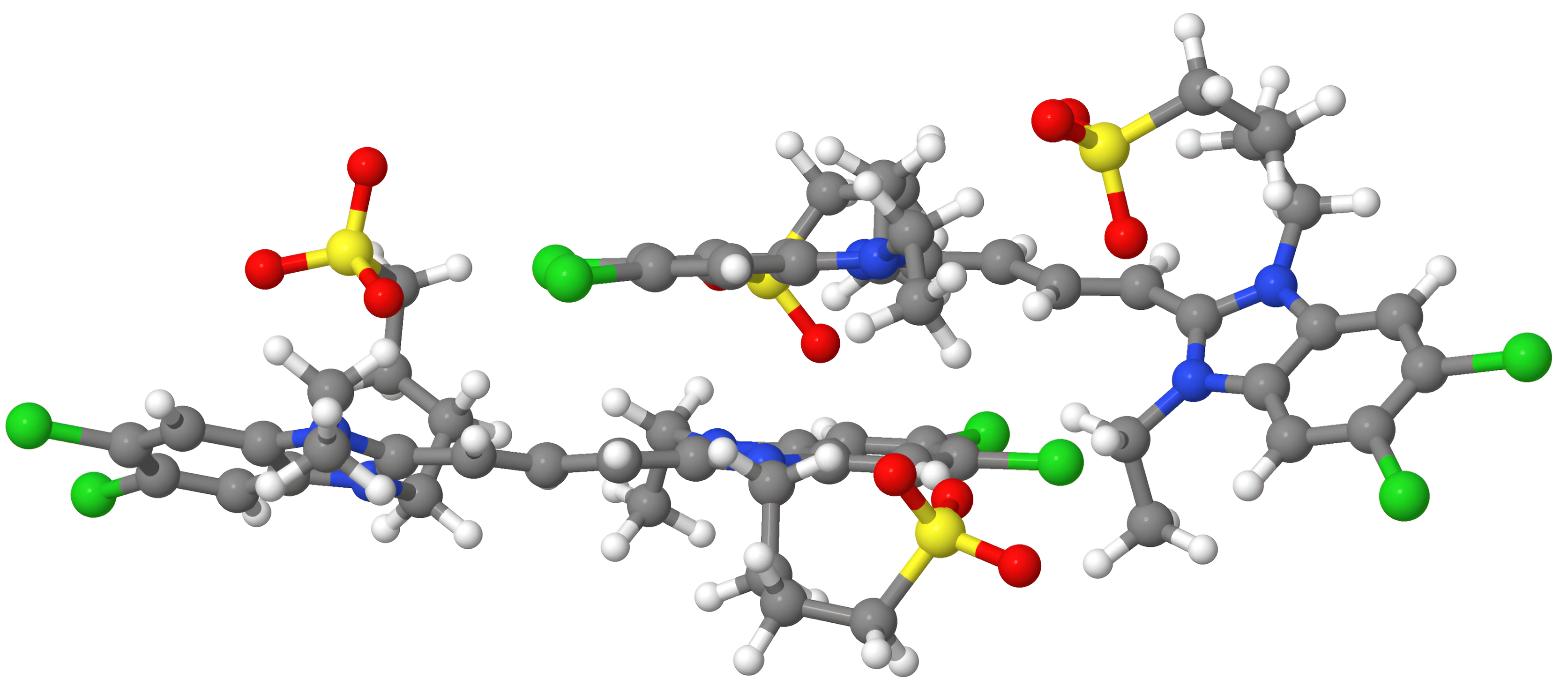}
    \label{fig:dimer_opt_side}
  \end{minipage}%
}\\[0.5em]

\subfloat[]{%
  \begin{minipage}[t]{\columnwidth}
    \centering
    \includegraphics[width=0.45\columnwidth]{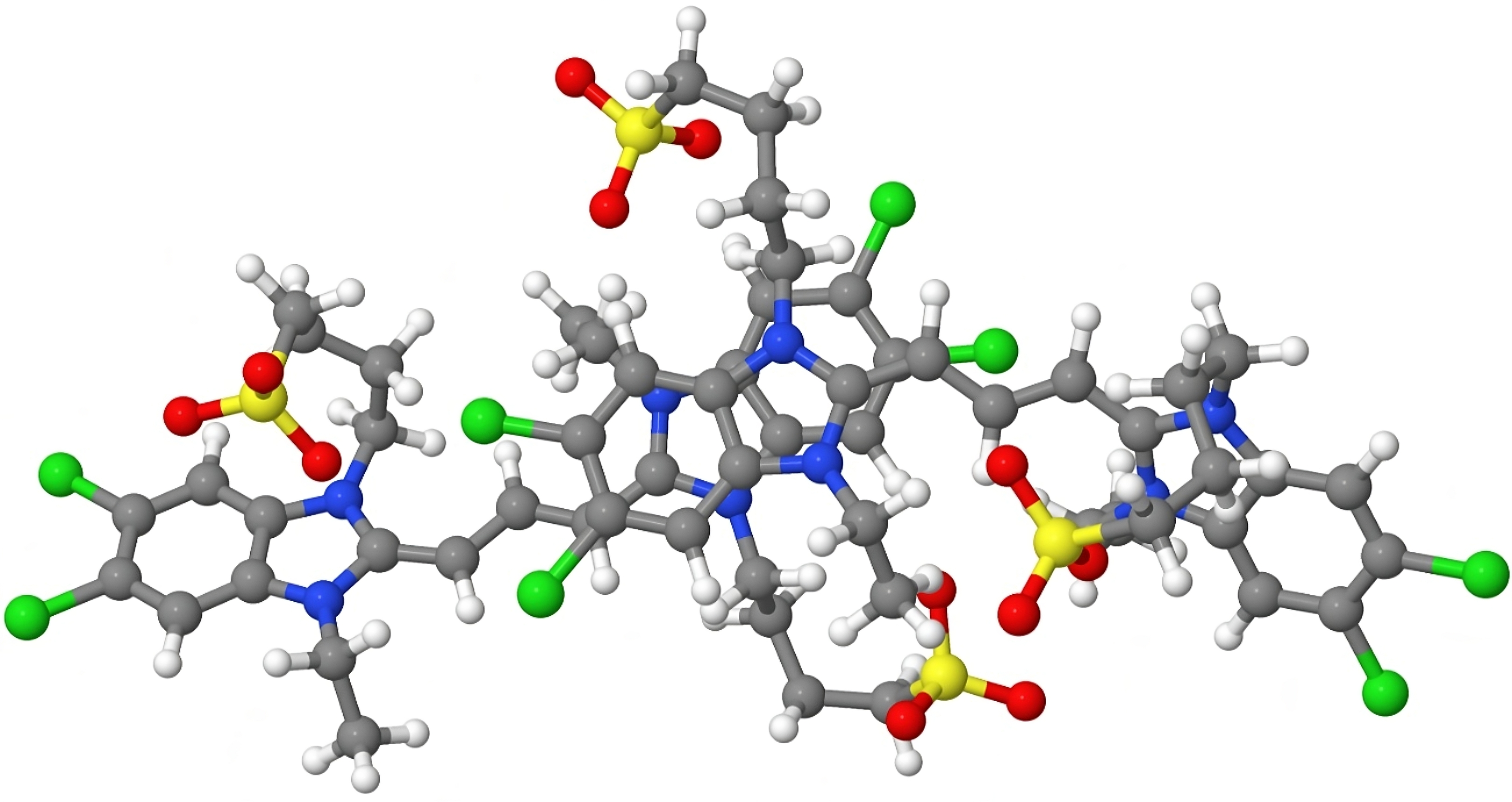}
    \label{fig:dimer_opt_front}
  \end{minipage}%
}

\caption{Geometry of the TDBC dimer structure in optimized conformation.
(a) side view; (b) front view.}
\label{fig:dimer_opt}
\end{figure}

% Dimers
Dimers were constructed by aligning the molecular transition dipoles at angles consistent with J-type aggregation, and subsequently optimized, following earlier work~\cite{coles2010characterization} (Fig. \ref{fig:dimer_opt}.
We observe again a curving of the aliphatic chains and an expected close intermolecular contact between $\pi$ systems. The dimer was further stabilized by approximately -22 kcal/mol per monomer of TDBC relative to the isolated monomer, at the electronic structure level. 
The two monomers in the optimized dimer exhibit dihedral angles of approximately 32$^\circ$ and 68$^\circ$, highlighting the heterogeneity of local torsions that can arise upon aggregation and that is reflected in the vibrational spectra.
These dimers are very helpful in understanding the behavior during aggregation, although they have to be taken with caution when considering quantitative agreement with real aggregates given that these consist of several (usually more than 10) molecules.

\begin{figure}[h!]
\centering

\subfloat[]{%
  \begin{minipage}[t]{\columnwidth}
    \centering
    \includegraphics[width=0.45\columnwidth]{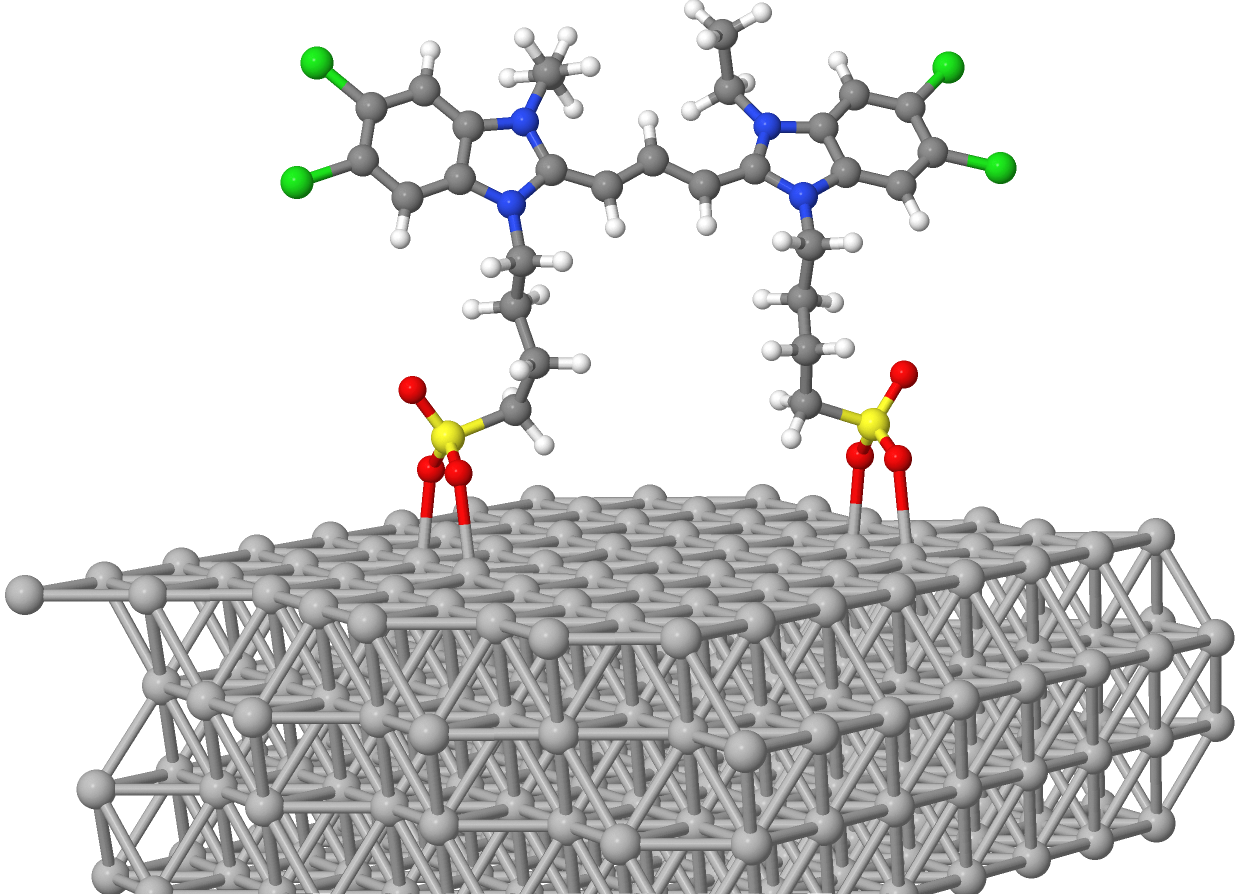}
    \label{fig:TDBC_Ag_up}
  \end{minipage}%
}\\[0.5em]

\subfloat[]{%
  \begin{minipage}[t]{\columnwidth}
    \centering
    \includegraphics[width=0.45\columnwidth]{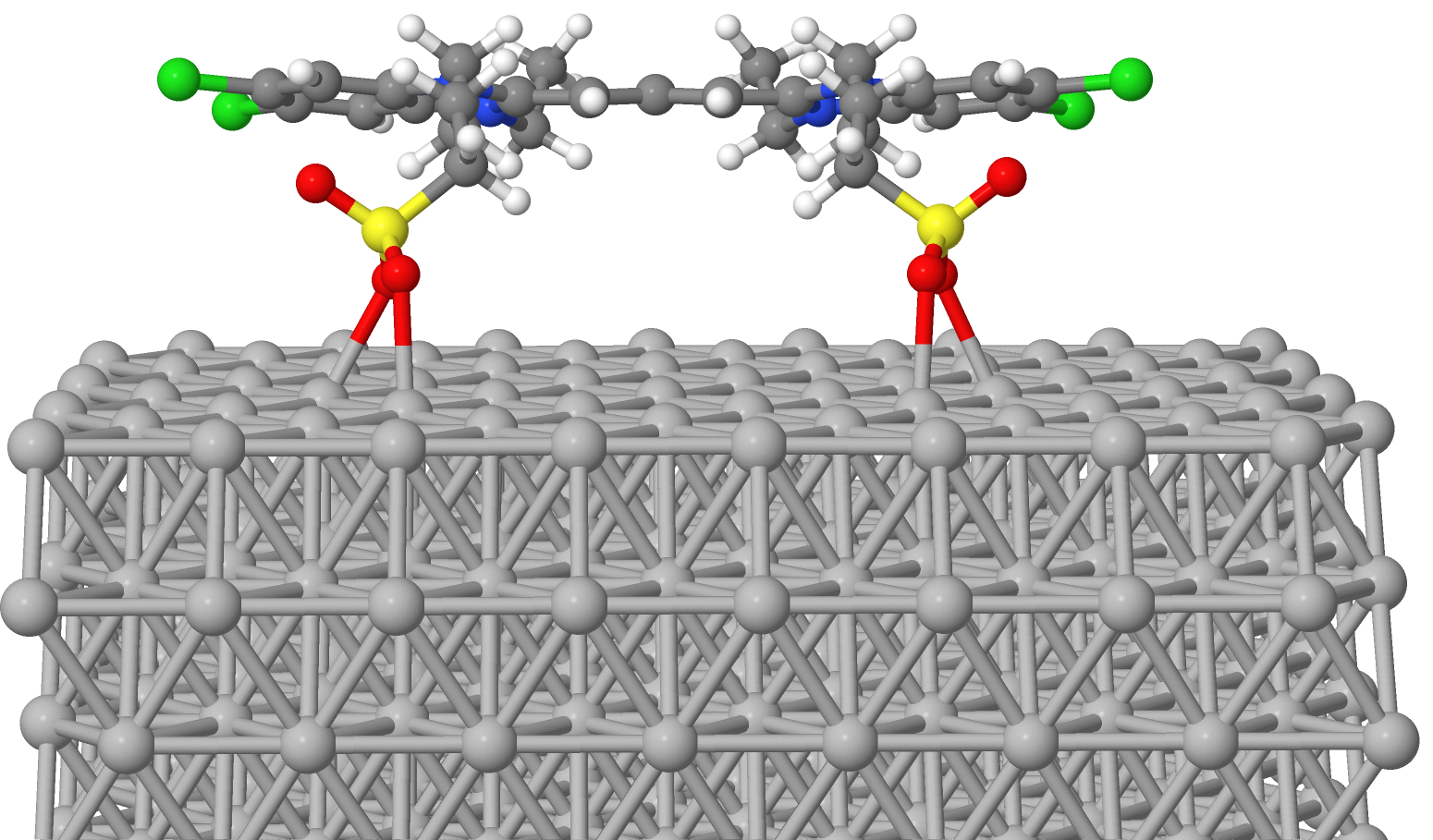}
    \label{fig:TDBC_Ag_flat}
  \end{minipage}%
}

\caption{Geometry of TDBC on a Ag surface.
(a) Molecule fully optimized; (b) molecule in the flat position.}
\label{fig:Ag}
\end{figure}

%TDBC on Ag
We also calculated the optimized geometry of TDBC on a Ag slab. The Ag slab atoms were kept fixed during all optimizations.
In a first calculation, the molecules were constrained to a flat geometry of the molecule (Fig. \ref{fig:Ag}.a), while for the second we allowed a full geometry optimization of the molecule (Fig. \ref{fig:Ag}.b). In both scenarios, the preferred adsorption moieties were the sulfonate groups via two bridging oxygen atoms. When the molecule was free to move, the most energetically favorable position was when the aromatic ring was parallel to the surface. This adsorption-induced planarization is consistent with the emergence of monomer-like Raman signatures and torsion-sensitive intensity ratios observed for plexcitons under blue-resonant excitation.
\newline

% Normal mode calculation
Normal modes were calculated for the monomeric and dimeric geometries and classified according to their dominant atomic motions (out-of-plane, in-plane, aliphatic chain) and spatial extent (localized on a single molecule or delocalized across two molecules). This classification provides the basis for interpreting aggregation- and adsorption-sensitive features in the Raman and THz-Raman spectra (see SI). 

Taken together, the DFT results establish a hierarchy of energetically accessible TDBC geometries that are consistent with the experimental constraints. The monomer exhibits a preferred twisted conformation, aggregation introduces heterogeneous torsional environments even at the dimer level, and adsorption on Ag favors planarization of the aromatic core via sulfonate binding. These trends provide a microscopic framework for interpreting the coexistence of aggregate-like and monomer-like vibrational signatures observed in plexcitonic assemblies.

\section{Discussion}

Below, we first discuss the geometry of TDBC J-aggregates in solution, then analyze how vibrational fingerprints reflect aggregation and disorder, and finally consider the most likely interfacial structures and their implications for plexciton design.
Our combined NMR, Raman, THz-Raman, and DFT study provides new insights into the structural organization of TDBC molecules in their monomeric, aggregated, and plexcitonic states. The main outcome is that while plexcitons retain many of the spectroscopic signatures of J-aggregates in water, their interfacial environment with Ag nanoparticles induces additional disorder and conformational freedom reminiscent of the monomeric dye. Furthermore, we detect a minority population of uncoupled monomers directly bound to the Ag surface. In addition to providing a basis for assigning the geometry of TDBC in various environments and structures, we also aim to identify the most important spectroscopic signatures that can be used when monitoring the geometry of similar cyanine dyes.  \newline

\textbf{Geometry of TDBC J-aggregates.}  
STM and fluorescence microscopy studies have established that cyanine and carbocyanine dyes adopt several packing motifs, including staircase, ladder, brickwork, and tubular arrangements \cite{crystalline_prokhorov_2014,prokhorov_polymorphism_2015}. For TDBC, two-dimensional sheet-like J-aggregates are the predominant structure in aqueous solution. Our NMR NOESY measurements further support this interpretation: intermolecular cross-peaks are most naturally compatible with an alternating, symmetric arrangement of sulfonate side chains rather than an asymmetric configuration. In particular, we suggest the NOESY cross-peaks between H$_{11}$ and H$_{13,14}$ as an indicator of alternating aggregation.   \newline

%\textbf{On the planarity of the aromatic system.}   

\textbf{Raman spectroscopic fingerprints of aggregation.}  
Low-frequency modes (below 900 cm$^{-1}$), which our DFT analysis shows are delocalized across multiple molecules, exhibit pronounced intensity enhancements in the J-aggregates and plexcitons. This finding echoes earlier reports of aggregation-enhanced Raman scattering (AERS), where collective out-of-plane modes become selectively amplified in J-aggregates \cite{coles2010characterization}. Coles et al.\ identified the 673 cm$^{-1}$ and 322 cm$^{-1}$ modes of TDBC as strongly enhanced upon aggregation. We also observe an unusually large 673 \cm mode. In addition we see the emergence above 500 \cm of a cluster of modes in the J-aggregates that are not present in the monomer (683, 730, 862, 1189, 1240, 1266 and 1344 \cm). These are also present in the plexcitons. 
The THz-region of aggregates shows different and more well-defined resonances in up to 400 \cm, providing enhanced sensitivity to long-range periodicity and collective packing. \newline

%Thus, the strongest indicators of fully formed aggregates are 671 \cm (AERS-enhanced), 176 \cm (long-range periodicity) and disappearance of monomeric bending modes below ~800 \cm.  

% \textbf{Uncertainties in the assignments.} \newline

\begin{figure}
    \centering
    \includegraphics[width=0.5\textwidth]{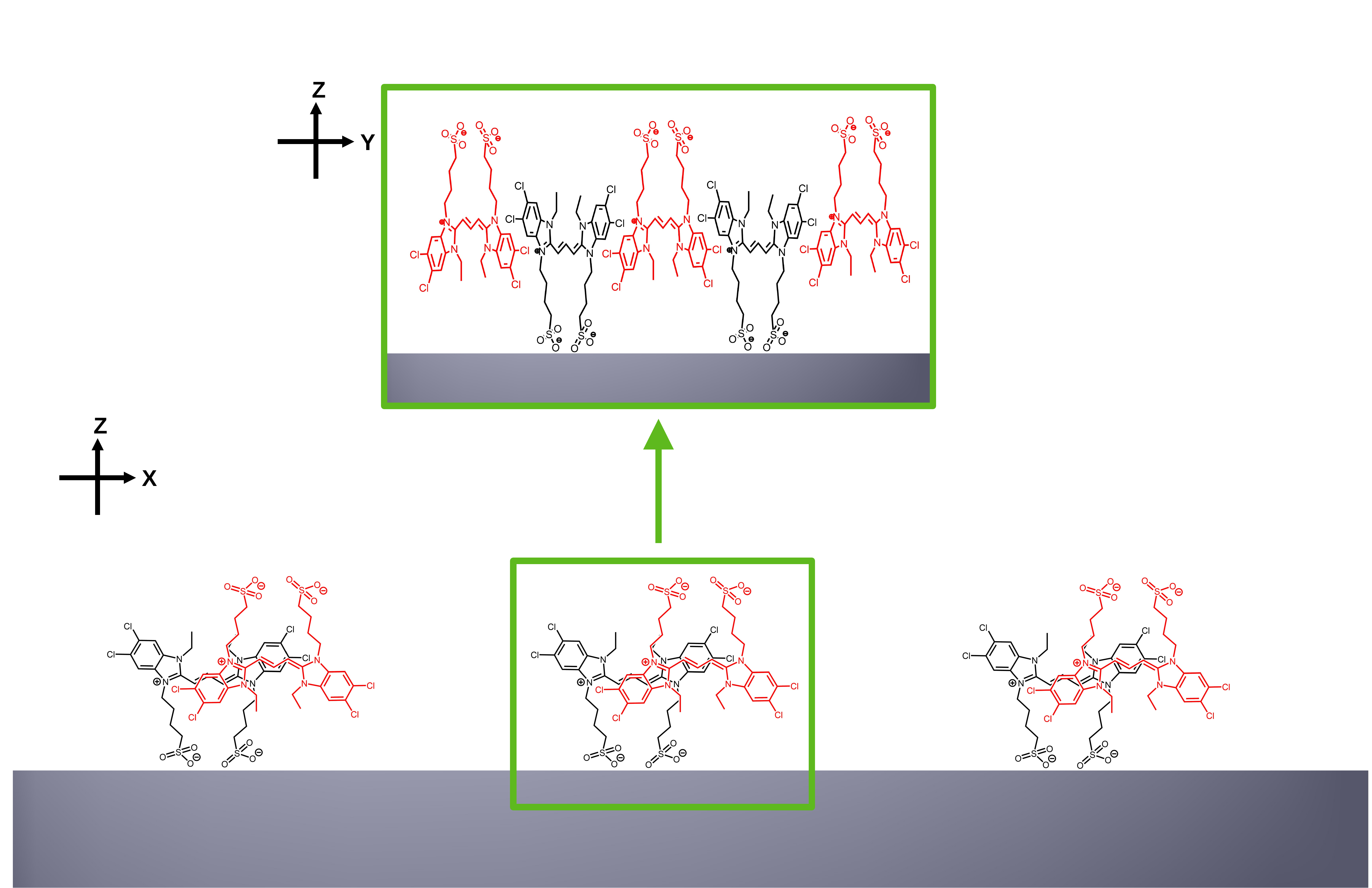}
    \caption{Inferred structure of TDBC on the Ag surface of the plasmonic nanoparticles.}
    \label{fig:most_likely_structure}
\end{figure}

\textbf{Most likely structures of TDBC adsorbed on Ag.}
While multiple interfacial geometries are in principle accessible, the combined energetic preference for sulfonate binding, Raman similarity to aggregates, and preservation of the aggregate emission frequency point toward the configuration shown in Fig. \ref{fig:most_likely_structure} as a plausible dominant motif. \newline

% \textbf{Conformational freedom and impurities.} TDBC is known to degrade and presents colorless impurities that appear in the NMR spectrum \cite{barotov_nearunity_2022}. We also identify the same impurities, which are more predominant in water than in methanol. A clear example is the appearance of three resonances corresponding to the methyl group, one of which is broadened and exhibiting NOESY cross-peaks with the sulfobutyl group, while the others do not. We have thus the possibility of having within the aggregate a fraction of monomer (also observed in the UVvis), and also the existence of photogenerated impurities. \newline

% This is no longer valid
%\textbf{Plexciton disorder and interfacial effects.}  
%When TDBC binds to Ag nanoprisms, the Raman and THz-Raman data indicate that the dyes maintain aggregated features, particularly in the low-frequency regime, while recovering monomer-like localized modes at higher frequencies {\color{red} This we have to remove}. This suggests a mixed state: lateral $\pi$--$\pi$ stacking interactions are preserved, but the anchoring sulfonate groups on the metal surface increase the heterogeneity of local environments, broadening the vibrational bands. An alternative explanation is that dangling sulfonate groups can adopt a conformation similar to that of the monomer. The broad THz-Raman feature around 125 cm$^{-1}$, absent in pure aggregates, exemplifies this interfacial disorder. Such structural variability is expected, as surface binding can disrupt the long-range periodicity seen in solution-based aggregates, as noted in earlier studies of dye--nanoparticle complexes \cite{Prokhorov2014}. \newline

\textbf{Comparison with encapsulated TDBC.}  
Interestingly, silica encapsulation of TDBC J-aggregates has recently been shown to ``freeze'' the brickwork geometry into rigid, defect-suppressed sheets, yielding near-unity quantum yields and exceptional robustness \cite{barotov_nearunity_2022,thanippuli_arachchi_bright_2024}. In contrast, our plexciton system demonstrates the opposite effect: attachment to a metallic surface introduces structural heterogeneity. This contrast highlights the competing roles of rigidifying (silica) versus supporting (metal) environments on the same molecular scaffold. The plexciton system therefore represents an interesting limit: strong exciton–plasmon coupling maintained despite locally broken 2D order. Indeed, a signature of strained TDBC monomer can also be seen in plexcitons (Figs. \ref{fig:plexciton_different_wavelengths_smooth}, \ref{fig:THz_Raman}).  \newline

\textbf{Implications for plexciton design.}  
Taken together, our results emphasize that plexciton performance is not solely determined by the number of coupled molecules nor by coupling strength alone, but also by their nanoscale geometry and the degree of structural disorder at the interface. The identification of vibrational fingerprints---most notably the Raman-active modes at $\sim$1600 \cm , $\sim$1489 cm$^{-1}$ and $\sim$793 cm$^{-1}$ among others---provides practical markers for diagnosing when dyes are in symmetric aggregates versus disordered surface-bound states. These insights open a path toward rational control of plexcitonic materials, for example by tuning surface functionalization or co-adsorbed ligands to balance aggregation strength and interfacial order.  \newline

From a theoretical perspective, the coexistence of aggregate-like and monomer-like geometries at the metal interface implies a distribution of excitonic energies and coupling strengths within the plexcitonic assembly. The structural heterogeneity identified here therefore provides microscopic context for models that treat plexcitons as disordered, open quantum systems with non-uniform light–matter coupling \cite{FinkelsteinShapiro2023}.

\section{Conclusion}

% spectrograph (Kymera 328i, Andor) 
Plexcitonic assemblies are promising platforms for the implementation of hybrid light-matter states in close contact to a chemical environment. Knowledge of the structure and its disorder is crucial to inform synthetic efforts of better materials, as well as to build better theoretical models to describe them. Through NMR and Raman spectroscopy supported by calculations, we have identified that in TDBC aggregates the sulfonate groups alternate above and below the aggregation plane. In plexcitons, this structure is preserved, although we observe the absence of longer range order as well as the existence of monomeric species. The dihedral angle between benzimidazole aromatic ring systems in the TDBC exists at an angle, with the surface binding tending to planarize the geometry, in particular for the monomers. In this investigation, we have unveiled key fingerprints of the different conformers, providing a benchmark for the synthesis of more complex plexcitons as well as establishing a framework for explaining the photophysics of TDBC-Ag model systems. 

\section{Supporting Information}

Supporting Information is available free of charge at the journal website.
Additional experimental details, supplementary Raman and THz-Raman spectra and extended NMR data. 

\section{Acknowledgements}

We thank Dr. Beatriz Quiroz García for technical assistance during the NMR experiments and Tomás Rocha-Rinza for useful discussions. 
D.F.S. acknowledges grant ECOS-CONAHCYT No. 321169 as well as funding from DGAPA-PAPIIT IA207223. M.L.P. acknowledges funding from CEA-Saclay and ANR. The work of Y.A.G.J. was supported by the UNAM Postdoctoral Program (POSDOC). G.P. acknowledge support from Dgapa-papiit IN104325. J.B.G. acknowledges a doctoral fellowship from SECIHTI.
J.H.S. acknowledges support from the Energy Area of Advance at Chalmers University of Technology.
This work benefited from the Resonance Raman platforms of I2BC, supported by the French Infrastructure for Integrated Structural Biology (FRISBI) ANR-10-INSB-05, and ECOS-Nord program (action M21P02). 
We acknowledge the use of the 700 MHz NMR instrument at the LURMN national laboratory.
Computations were performed at NSC Tetralith provided by the National Academic Infrastructure for Supercomputing in Sweden (NAISS), partly funded by the Swedish Research Council (grant no. 2022 -06725).

\bibliography{sbc,plexciton_all,Aggregates_structure,RR_aggregate_structure}

\clearpage

\newpage

\end{document}